\documentclass[reprint,aps,prx,superscriptaddress]{revtex4-2}
\usepackage[final]{graphicx}
\usepackage[utf8]{inputenc}
\usepackage[top=0.6in, bottom=1in, left=0.8in, right=0.8in]{geometry}
\usepackage[hidelinks]{hyperref}
\usepackage[normalem]{ulem}
\usepackage{bm,physics,amsmath,amssymb,xcolor}

\graphicspath{{./figures/}}

\begin{document}

\title{Theory of angular momentum transfer from light to molecules}

\author{Mikhail Maslov}
\email{mikhail.maslov@ist.ac.at}
\affiliation{Institute of Science and Technology Austria (ISTA), Am Campus 1, 3400 Klosterneuburg, Austria}
\author{Georgios M. Koutentakis}
\email{georgios.koutentakis@ist.ac.at}
\affiliation{Institute of Science and Technology Austria (ISTA), Am Campus 1, 3400 Klosterneuburg, Austria}
\author{Mateja Hrast}
\affiliation{Institute of Science and Technology Austria (ISTA), Am Campus 1, 3400 Klosterneuburg, Austria}
\author{Oliver H. Heckl}
\affiliation{Christian Doppler Laboratory for Mid-IR Spectroscopy and Semiconductor Optics,
Faculty Center for Nano Structure Research,
Faculty of Physics, University of Vienna,
Boltzmanngasse 5, 1090 Vienna, Austria}
\author{Mikhail Lemeshko}
\email{mikhail.lemeshko@ist.ac.at}
\affiliation{Institute of Science and Technology Austria (ISTA), Am Campus 1, 3400 Klosterneuburg, Austria}

\begin{abstract}
We present a theory describing interaction of structured light, such as light carrying orbital angular momentum, with molecules. The light-matter interaction Hamiltonian we derive is expressed through  couplings between spherical gradients of the electric field and the (transition) multipole moments of a particle of any non-trivial rotation point group. Our model  can therefore accommodate for an arbitrary complexity of the molecular and electric field structure, and can be straightforwardly extended to atoms or nanostructures. Applying this framework to  ro-vibrational spectroscopy of molecules, we uncover the general mechanism of angular momentum exchange between the spin and orbital angular momenta of light, molecular rotation and its center-of-mass motion. We show that the non-zero vorticity of Laguerre-Gaussian beams can strongly enhance certain ro-vibrational transitions that are considered forbidden in the case of non-helical light. We discuss the experimental requirements for the observation of these forbidden transitions in state-of-the-art spatially-resolved spectroscopy measurements.
\end{abstract}

% Another version:
%The interaction of molecules with the orbital angular momentum of light has long been argued to benefit structural studies and quantum control of molecular ensembles. We derive a general description of the light-matter interaction in terms of the coupling between spherical gradients of the electric field and an effective molecular charge density that exactly reproduces molecular multipole moments. Our model can accommodate for an arbitrary complexity of the molecular structure and is applicable to any electric field, with the exception of tightly focused beams. Within this framework, we derive the general mechanism of angular momentum exchange between the spin and orbital angular momenta of light, molecular rotation and its center-of-mass motion. We demonstrate that vortex beams strongly enhance certain ro-vibrational transitions that are considered forbidden in the case of a non-helical light. Finally, we discuss the experimental requirements for the observation of novel transitions in state-of-the-art spatially-resolved spectroscopy measurements.

\maketitle

\section{Introduction}
\label{sec_intro}

Light-matter interactions are at the heart of many  disciplines, from imaging~\cite{Balasubramanian_review_2023} and spectroscopy~\cite{Herzberg_book,Corney_book} to quantum control~\cite{Shapiro_book} and sensing~\cite{Degen_review_2017}, to quantum information~\cite{Monroe_review_2002} and ultracold chemistry~\cite{Liu_review_2022}. In many of these systems, rotations of some sort -- whether of molecules, spins, or light polarization -- play a  major role in defining the emergent physics. While being an elementary class of spatial transformations, they form a  non-Abelian group~\cite{Wigner_book,Broecker_2013}, which makes them promising for foundational studies of topological physics~\cite{Karle_2023,Albert_2024} and the design of quantum computing codes~\cite{Albert_2020}.

Rotations are directly associated with the concept of angular momentum (AM). The spin, or intrinsic, AM of light has been extensively studied in the \mbox{past~\cite{Beth_1936,Bliokh_OAM,Andrews_Book}.} However, only a couple decades ago, it was shown that, in addition to the spin, light can possess the extrinsic, or orbital, angular momentum (OAM) in the form of a helical beam phase~\cite{Allen_1992,Bliokh_OAM,Andrews_Book}. So-called twisted photons with large values of  OAM can be generated by combining fundamental laser modes~\cite{Tamm_1990,Allen_1992}, i.e., using a diffraction grating~\cite{Bazhenov_1992}, a spiral phase plate~\cite{Schemmel_2014} or a metasurface~\cite{Karimi_2014}. Outside the regime, where the spin and OAM are strongly coupled~\cite{Zhao_2007,Bliokh_SOC}, they represent two conceptually different characteristics of light. The former is related to the polarization of the electric field, while the latter is induced by its spatial gradient~\cite{Bliokh_OAM,Andrews_Book}.

Here, motivated by the multifaceted nature of the AM of light, as well as by the ubiquity of light-matter interactions, we introduce a general theoretical framework of angular momentum exchange between an optical field and a particle. To narrow down our explanation and make it illustratory, we choose the ro-vibrational spectroscopy of molecules as the prototypical system of interest. Unlike in other scenarios, where the interaction with an optical field induces interwoven relaxation subprocesses that are hard to trace, ro-vibrational spectroscopy can be described with sufficient accuracy involving relatively few  rotational degrees of freedom. Moreover, the emergent angular momentum exchange is characterized by selection rules, which encapsulate the essential physical behavior and are accessible through a direct experimental measurement.

Our theoretical approach is conceptually similar to the Wigner-Eckart theorem~\cite{Hall_Lie}. We suggest addressing the general problem of the light-matter interaction via an analytical framework that is designed around the concept and algebra of angular momentum. This way most of the interaction's complexity can be absorbed by the coefficients of the resulting spherical expansion, similar to the reduced matrix elements of the Wigner-Eckart decomposition. 

Previous studies on ro-vibrational spectroscopy using twisted light~\cite{Babiker_2002,Alexandrescu_2006,Mondal_2014,Mukherjee_2018} attempted to track the full problem of the light-matter interaction to get access to the selection rules for the AM exchange. As a result, they are only applicable to a narrow class of systems. In particular, they are restricted to specific point-charge models or electric field profiles. In contrast, our approach circumvents the detailed analysis of the molecular or optical field structure, replacing it with the analysis of the coupling between electric field gradients and molecular multipole moments. The resulting selection rules depend only on whether the spherical expansion coefficients are vanishing or not, providing a direct connection of AM exchange to the molecular and optical field symmetry.

As in the case of the Wigner-Eckart theorem, the transition rates within our theory are interrelated. This creates a possibility to calculate rates of experimentally unaccessible transitions, by linking them to their respective measurable counterparts. This feature of our theory might be important for future experimental developments, e.g., in metrology. It allows bypassing complex ab-initio computations that are prone to inaccuracies due to simplifying assumptions, stemming, for example, from the Born-Oppenheimer approximation, relativistic effects or spin-orbit coupling.

To benchmark our theory and demonstrate its predictive power, we use the suggested framework to derive the selection rules for the ro-vibrational spectroscopy using Laguerre-Gaussian beams. We reveal that the helicity of light (OAM) can substantially enhance rotational transitions that are considered forbidden in the case of a spin-only light. Notably, instead of merely estimating amplitudes of the enhanced transitions, our theory allows us to determine the origin of the enhancement. Finally, we suggest a possible experimental scheme that may be capable of verifying our findings, by measuring the ro-vibrational spectrum of gas phase molecules.

The paper is structured as follows. In Sec.~\ref{sec_multipole_moments}, we introduce the effective model for the multipole moments. On the example of a molecule, we explain how internal degrees of freedom, like vibrations and rotations, can be embedded into the model. Our general idea can be applied to other scenarios of the light-matter interaction, as explained in App.~\ref{app_extensions}. In Sec.~\ref{sec_hamiltonian}, we derive the general expression for the interaction of an optical field with multipoles. Our derivation is designed to highlight the angular momentum exchange and is valid for \textit{any rotating particle and any profile of the electric field}. Motivated by the ro-vibrational spectroscopy, we derive the general selection rules for the relevant optical fields. In Sec.~\ref{sec_LG_beams}, we consider Laguerre-Gaussian beams and study the impact of electric field gradients on the OAM transfer to the molecular (internal) rotation. We infer scenarios in which vortex beams offer an advantage over the non-twisted light.  In Sec.~\ref{sec_transition_rates}, we calculate transition rates of ro-vibrational transitions. In Sec.~\ref{sec_experiment}, we suggest a prototypical experimental scheme that might be capable of measuring the changes to the ro-vibrational spectrum, induced by the OAM of light. We conclude the paper with the summary of possible extensions and applications of our theory in Sec.~\ref{sec_conclusion}.

\section{Molecular multipole moments}
\label{sec_multipole_moments}

The principal analytical tool for studying interaction of particles with optical fields is the multipole expansion~\cite{Jackson,Power_book,Fiutak_1963}. Within this paradigm, the electromagnetic field is considered to interact with the \textit{multipole moments} of the entire particle rather than its individual constituents, like nuclei and electrons in a molecule. In the case of the ro-vibrational spectroscopy, the accuracy of this expansion is assured by the immense separation of particle and field length scales. For instance, for the carbon monosulfide (CS) molecule, the wavelength corresponding to the lowest vibrational transition $(\nu=0\to1)$ is $\lambda_{\text{vib}}\approx8\,\text{µm}$, while the bond length is $a\approx1.5\,\text{\r{A}}$~\cite{NIST}. With the appropriate calculation of multipole moments, the multipole expansion can be exact, as demonstrated in Ref.~\cite{Alaee_2018} for Mie scattering. The multipole expansion does not eliminate any complexity from the general problem of the light-matter interaction. Instead, it provides a simple analytical framework for the interaction of the field with multipoles, which embed the internal structure of the particle.

Motivated by ro-vibrational spectroscopy, we consider a nonmagnetic molecule and define the dependence of its electric multipole moments on the rotation and internal vibrations.  To keep our model simple, we omit other field-induced displacements of intramolecular charges, like electronic transitions. Although we focus on the particular case of molecules, our idea applies to small particles in general. Following our example, one can add other degrees of freedom to the molecular model or adapt the model to describe nanostructures or electronic transitions in atoms (see App.~\ref{app_extensions}). Notably, our model does not aim at replacing state-of-the-art quantum chemistry methods for calculating multipole moments. It rather provides a simple analytical formulation that is built on top of these ab-initio calculations.

\begin{figure}[p]
	\centering\noindent
    \includegraphics[width=\linewidth]{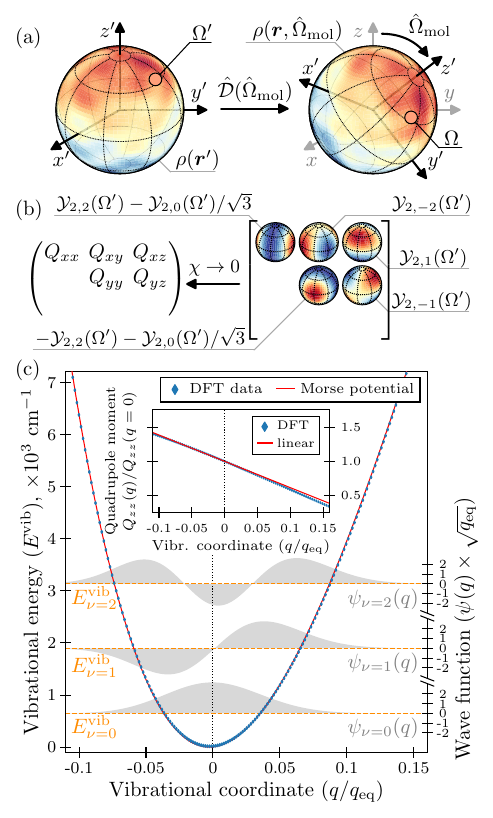}
    \caption{Outline of the effective model for molecular multipole moments. (a)~At equilibrium, the molecular point group~$\mathcal{G}_\text{mol}$ is described by spherical multipole moments~$\alpha_{\lambda,\mu}$ that can be obtained experimentally or via \textit{ab-initio} calculations. We encode these multipole moments into a charge distribution $\rho(\bm{r}')$ on a sphere \mbox{with a radius $\chi\to 0$} in the molecular frame~(left). The corresponding laboratory-frame charge distribution $\rho(\bm{r},\hat{\Omega}_{\text{mol}})$, which describes the molecular rotation, can be obtained using the rotation operator~$\hat{\mathcal{D}}(\hat{\Omega}_{\text{mol}})$, where $\hat{\Omega}_{\text{mol}}$ is the molecular orientation operator~(right). (b)~The spherical multipole moments $\alpha_{\lambda,\mu}$ can be directly mapped to axial multipole moments, as exemplified here by the quadrupole moment matrix $Q_{i,j}$, where ${i,j\in\{x,y,z\}}$. (c)~The quadrupole moment matrix~$Q_{i,j}^{\nu,\nu'}$, associated with the vibrational transition $\nu\to\nu^{\prime}$, can be calculated numerically using DFT. For the CS molecule, we obtain the adiabatic potential energy curve~$E^{\text{vib}}(q)$~\mbox{(markers)} and the associated eigenenergies $E^\text{vib}_\nu$~(dashed lines) and eigenstates $\psi_\nu(q)$~(shades), where $q$ is the vibrational coordinate. Both the spectrum and eigenstates are asymmetric with respect to the equilibrium internuclear distance $q_\text{eq}$. (inset)~The coordinate dependence of the quadrupole moment $Q_{zz}(q)$. To obtain transition quadrupole moments, we calculate the expectation values of $Q_{zz}(q)$ with respect to different eigenstates $\psi_{\nu(\nu')}(q)$~(see Eq.~\eqref{eq_transition_moment}).}
    \label{fig_charge_density}
\end{figure}

We begin by considering a body-fixed reference frame, which we herewith call the \textit{molecular frame}. In particular, we consider the coordinate axes, co-aligned with the molecular orientation $\hat{\Omega}_\text{mol}$ and with the origin at its center of mass. 
The orientation of the molecule is uniquely defined in a way that respects symmetries of the molecular point group $\mathcal{G}_\text{mol}$. The mathematical apparatus behind this process is summarized in App.~\ref{app_molecular_rotation_group}. In the molecular frame, we suggest an effective model for the molecular multipole moments. In particular, we introduce a charge distribution $\rho(\bm{r}')$ on a sphere of an infinitesimally small radius $\chi\to 0$, centered at the molecular frame origin (see Fig.~\ref{fig_charge_density}(a)). We expand $\rho(\bm{r}')$ in terms of the real-valued spherical harmonics
\begin{equation}
    \rho(\bm{r}')=\sum\limits_{\lambda,\mu}\alpha_{\lambda,\mu}\mathcal{Y}_{\lambda,\mu}(\Omega')\delta(r'-\chi)/\chi^{\lambda+2}\,,
    \label{eq_charge_density_mol}
\end{equation}
where $\lambda = 1, 2, \dots$, $|\mu| \le \lambda$, $\bm{r}'=\{r',\theta',\phi'\}=\{r',\Omega'\}$ are the molecular-frame spherical coordinates 
and the real-valued harmonics $\mathcal{Y}_{\lambda,\mu}(\Omega')$ are defined as
\begin{equation}
    \mathcal{Y}_{\lambda,\mu}(\Omega')=\begin{cases}
    i\big(Y_{\lambda,\mu}(\Omega')-Y^\ast_{\lambda,\mu}(\Omega')\big)/\sqrt{2}&\mu<0\\
    Y_{\lambda,0}(\Omega')&\mu=0\\
    \big(Y_{\lambda,-\mu}(\Omega')+Y^\ast_{\lambda,-\mu}(\Omega')\big)/\sqrt{2}&\mu>0\end{cases}\,,
    \label{eq_real_harmonics}
\end{equation}
where $Y_{\lambda,\mu}(\Omega^\prime)$ are the complex-valued spherical harmonics with the Condon-Shortley phase~\cite{Varshalovich}. 

The choice of the charge distribution~\eqref{eq_charge_density_mol} is not accidental. Harmonics $\mathcal{Y}_{\lambda,\mu}(\Omega')$ form the complete basis on the sphere, parameterized by the angle $\Omega'$. The expansion coefficients~$\alpha_{\lambda,\mu}$, called \textit{spherical} multipole moments~\cite{Jackson}, fully reflect the structure of the molecular \mbox{(symmetry)} point group $\mathcal{G}_\text{mol}$, as discussed in Ref.~\cite{Gelessus_1995}. For instance, for a heteronuclear diatomic molecule, like CS: $\alpha_{\lambda,\mu\neq0}=0$. Note that the $\lambda = 0$ term is excluded from Eq.~\eqref{eq_charge_density_mol}, as we restrict our analysis to the particles of zero net charge. In an experiment, it is practical to define \textit{axial} multipole moments, like the Cartesian dipole moment components $d_x$, $d_y$ and~$d_z$. They can be mapped to $\alpha_{\lambda,\mu}$ by integrating the charge distribution $\rho(\bm{r}')$ with the Cartesian tensor of a rank~$\lambda$. For instance, for the given dipole moment vector ${\bm{d}=\{d_x,d_y,d_z\}}$, the condition
\begin{equation}
    \begin{pmatrix}d_x\\d_y\\d_z\end{pmatrix}=
    \lim_{\chi\to 0}\int\rho(\bm{r}')
    \begin{pmatrix}x'\equiv r'\sin{\theta'}\cos{\phi'}\\y'\equiv r'\sin{\theta'}\sin{\phi'}\\z'\equiv r'\cos{\theta'}\end{pmatrix}\mathrm{d}^3\bm{r}'\,,
    \label{eq_dipole_integral}
\end{equation}
yields: $\{\alpha_{1,-1},\alpha_{1,0},\alpha_{1,1}\}=\sqrt{\frac{3}{4\pi}}\{d_y,d_z,d_x\}$. Similar mappings can be derived for all higher-order multipole moment tensors,~e.g.,~quadrupole, as shown in Fig.~\ref{fig_charge_density}(b). When calculating the corresponding integrals, which are similar to Eq.~\eqref{eq_dipole_integral}, the denominator $1/\chi^{\lambda+2}$ in Eq.~\eqref{eq_charge_density_mol} balances out the $\chi$ contributions stemming from Cartesian tensor and Jacobian, assuring that the multipole moments are finite in the limit $\chi\to0$.

To transform the effective charge distribution $\rho(\bm{r}')$ into the laboratory frame distribution $\rho(\bm{r},\hat{\Omega}_\text{mol})$, one needs to cast the molecular-frame spherical harmonics (of the angle $\Omega'$) in terms of the laboratory-frame angle~$\Omega$, using the rotation rule 
\begin{align}
    Y_{\lambda,\mu}(\Omega')=\,&\hat{\mathcal{D}}(\hat{\Omega}_\text{mol})Y_{\lambda,\mu}(\Omega)\nonumber\\
    &=\sum_{\zeta}D^{\lambda}_{\zeta,\mu}(\hat{\Omega}_\text{mol})Y_{\lambda,\zeta}(\Omega)\,,
    \label{eq_sph_harm_rotation}
\end{align}
where $\hat{\mathcal{D}}(\hat{\Omega}_\text{mol})$ is the rotation operator that connects the molecular and laboratory frames~(see Fig.~\ref{fig_charge_density}(a)) and $D^{\lambda}_{\zeta,\mu}(\hat{\Omega}_\text{mol})$ are the irreducible representations of the molecular rotation group (see App.~\ref{app_molecular_rotation_group}). Provided that the spherical multipole moments $\alpha_{\lambda,\mu}$ already embed the symmetries of the molecular point group $\mathcal{G}_\text{mol}$, when applying the transform to the distribution~\eqref{eq_charge_density_mol}, one can simply use Wigner D-matrices~\cite{Varshalovich} in Eq.~\eqref{eq_sph_harm_rotation}.

%%%%%%%%%%%%%%%%%%%%%%%%%%%%%%%%%%%%%%%%%%%%%%%%%%
%%%%%%%%%%%%%%%%%%%%%%%%%%%%%%%%%%%%%%%%%%%%%%%%%%

In addition to reflecting the point group $\mathcal{G}_\text{mol}$~\cite{Gelessus_1995}, multipole moments also describe the longitudinal geometry of the molecule~\cite{Green_2007}, including its vibrations. In particular, numerous studies have shown that the dipole~\cite{Winnewisser_1968,Patel_1979} and quadrupole~\cite{Truhlar_1972} moments are different in the ground and excited vibrational states. The vibrational dependence of multipoles is difficult to formalize in the case of an arbitrary polyatomic molecule. One requires an accurate model of the coupling between molecular rotations and different vibrational excitations, like the Watson Hamiltonian~\cite{Watson_1968,Watson_1993}, discussed in App.~\ref{app_extensions}. Apart from complicating the definition of the molecular orientation~$\hat{\Omega}_{\text{mol}}$ (see App.~\ref{app_molecular_rotation_group}), the ro-vibrational coupling effectively mixes light-induced excitations of the molecule, which severely complicates the analysis of the AM exchange, as explained in Sec.~\ref{sec_hamiltonian}. For this reason, we restrict our focus only to certain molecules. In particular, we assume that symmetries of the molecule in all vibrationally excited states form the same point group $\mathcal{G}_\text{mol}$ as in the ground state, which makes it possible to decouple rotations and vibrations. Nonetheless, as discussed in Sec.~\ref{sec_experiment}, we still account for the shift in ro-vibrational energies due to this coupling.

After separating vibrations from rotations, vibrational transition multipole moments in the molecular frame can be obtained using numerical \textit{ab-initio} methods~\cite{Pulay_1979,Pulay_1981}. In Fig.~\ref{fig_charge_density}(c), we present our calculation for the CS molecule, which has a single vibrational mode. Our analysis is based on the StoBe-deMon implementation~\cite{Stobe_2014} of density functional theory~\cite{Jones_2015}. By using the local Perdew-Wang exchange-correlation potential of Ref.~\cite{Perdew_1992} and constraining the Kohn-Sham orbitals~\cite{Kohn_Sham_1965, Jones_2015} to be rotationally invariant around the interatomic axis, we obtain the adiabatic potential energy curve $E^{\text{vib}}(q)$ (markers). Here the vibrational coordinate $q$ is the shift of the internuclear distance from its equilibrium value $q_{\text{eq}}\approx 1.552\,\text{\AA}$. The resulting curve can be reasonably approximated by the anharmonic Morse potential~(solid line). We also calculate the vibrational levels $E^{\text{vib}}_\nu$~(dashed lines) and the corresponding wave functions $\psi_{\nu}(q)$~(shades), which are asymmetric with respect to the equilibrium coordinate~$q_{\text{eq}}$. In the inset of Fig.~\ref{fig_charge_density}(c), we plot the dependence of the quadrupole moment $Q_{zz}$ on the vibrational coordinate $q$~(markers), which has non-linear profile~(cf.~solid line). The transition quadrupole moment matrix can be straightforwardly obtained by calculating the integral
\begin{equation}
    Q_{i,j}^{\nu,\nu'}=\int\psi^{\ast}_{\nu'}(q)Q_{i,j}(q)\psi_{\nu}(q)\mathrm{d}q\,,
    \label{eq_transition_moment}
\end{equation}
where $i,j\in\{x,y,z\}$. 

Finally, we summarize our model using the particular example: a heteronuclear diatomic molecule, \mbox{like CS.} Due to the U$_1$ symmetry of such a molecule, the dipole moment vector $\bm{d}=\{0,0,\hat{d}^{\text{vib}}_z\}$ and quadrupole moment matrix $Q_{i,j}=-\hat{Q}^{\text{vib}}_{z,z}\,\text{diag}\{1/2,1/2,-1\}$, where $\hat{d}^{\text{vib}}_z$ and $\hat{Q}^{\text{vib}}_{z,z}$ are operators with respect to molecular vibrational states. Omitting all the higher multipole moments, the effective charge distribution reads
\begin{align}
    &\rho_{\text{diatomic}}\big(\bm{r},\hat{d}^{\text{vib}}_z,\hat{Q}^{\text{vib}}_{zz},\hat{\Omega}_{\text{mol}}\big)\nonumber\\
    &=\sum\limits_{\mu}\hat{d}^{\text{vib}}_z Y_{1,\mu}\big(\hat{\Omega}_{\text{mol}}\big)Y_{1,-\mu}(\Omega)\frac{\delta(r-\chi)}{\chi^3}\nonumber\\
    &+\sum\limits_{\mu}\frac{3}{2}\hat{Q}^{\text{vib}}_{z,z} Y_{2,\mu}\big(\hat{\Omega}_{\text{mol}}\big)Y_{2,-\mu}(\Omega)\frac{\delta(r-\chi)}{\chi^4}\,.
    \label{eq_charge_density_diatomic}
\end{align}
Note that multipole moment tensors $\bm{d}$ and $Q_{i,j}$ have a more complicated structure for polyatomic (non-linear) molecules. Nonetheless, provided that the analytical or numerical dependence of multipole moment tensors on internal degrees of freedom is known, our multipole model of the light-matter interaction is valid for \textit{any molecular geometry}.

\section{Light-matter interaction and angular momentum exchange}
\label{sec_hamiltonian}

After specifying how the internal degrees of freedom of a particle can be embedded into the definition of its multipole moments, we define the Hamiltonian of the interaction between the optical field and multipoles.

Our starting point is the non-relativistic multipolar QED Hamiltonian, which can be obtained from the Coulomb gauge Hamiltonian by performing the unitary Power-Zienau-Wooley (PZW) transformation~\mbox{\cite{Power_1959,Atkins_1970,Wooley_1971}}. Both the aforementioned Hamiltonians are fully identical for the on-energy-shell processes, while PZW Hamiltonian gives, arguably, a better representation for the off-energy-shell processes~\cite{Power_1959,Andrews_2018}. We consider a spinless particle, thus we can safely omit its interaction with the magnetic field further on. The resulting Hamiltonian reads: $\mathcal{H}=\mathcal{H}_0+\mathcal{H}_{\text{int}}$, where the light-matter interaction can be expressed as
\begin{equation}
    \mathcal{H}_{\text{int}}=-\int\big(\bm{\mathcal{P}}(\bm{r}_\text{EF})\cdot\bm{E}(\bm{r}_\text{EF})\big)\mathrm{d}^3\bm{r}_\text{EF}+\text{H.c.}\,,
    \label{eq_hamiltonian}
\end{equation}
where $\bm{r}_\text{EF}$ is a coordinate in the laboratory frame. The Hamiltonian $\mathcal{H}_0$ describes the ``unperturbed'' system,~i.e.,~the electric field and particle that do not interact. In the case of a molecule, this term includes the kinetic energy of the center of mass, rotational and vibrational molecular energies, which are discussed in Sec.~\ref{sec_experiment}, as well as the energy of the electric field
\begin{equation}
    \mathcal{E}_{\text{EF}}=\frac{\varepsilon_0}{2}\int|\bm{E}(\bm{r}_\text{EF})|^2\mathrm{d}^3\bm{r}_\text{EF}\,,
    \label{eq_EF_energy}
\end{equation}
where $\varepsilon_0$ is the vacuum permittivity. Following the common practice, we also put the infinite self-energy term $\int|\bm{\mathcal{P}}(\bm{r}_\text{EF})|^2\mathrm{d}^3\bm{r}_\text{EF}$ into $\mathcal{H}_0$. The divergence of this integral, its origin and remedy are discussed in Ref.~\cite{Andrews_2018}. 

To define the polarization field $\bm{\mathcal{P}}(\bm{r}_\text{EF})$, we begin with the polarization $\bm{\mathcal{P}}_{\text{PZW}}(\bm{r}_\text{EF})$ of the PWZ Hamiltonian. It provides the exact description of the generic system of $N$ point-charges
\begin{equation}
    \bm{\mathcal{P}}_{\text{PZW}}(\bm{r}_\text{EF})=\sum\limits_{i=1}^{N} q_{i}\bm{r}_{i}\int\limits_{0}^{1}\mathrm{d}\eta\,\delta\big(\bm{r}_\text{EF}-\bm{R}-\eta\bm{r}_i\big)\,,
    \label{eq_PWZ_point_charges}
\end{equation}
where $q_i$ and $\bm{r}_i$ are the charges and their coordinates in the center-of-mass reference frame, and $\bm{R}$ is the center-of-mass position in the laboratory frame. This relation can be straightforwardly generalized to describe the polarization field $\bm{\mathcal{P}}(\bm{r}_\text{EF})$, generated by the continuous charge distribution $\rho(\bm{r})$, introduced in Sec.~\ref{sec_multipole_moments}
\begin{equation}
    \bm{\mathcal{P}}(\bm{r}_\text{EF})=\lim_{\chi\to 0}\int\limits\mathrm{d}^3\bm{r}\,\rho(\bm{r})\bm{r}\int\limits_{0}^{1}\mathrm{d}\eta\,\delta\big(\bm{r}_\text{EF}-\bm{R}-\eta\bm{r}\big)\,.
    \label{eq_PZW_density}
\end{equation}
Substituting the polarization field~\eqref{eq_PZW_density} into Eq.~\eqref{eq_hamiltonian}, we obtain
\begin{equation}
    \mathcal{H}_\text{int}=-\lim_{\chi\to 0}\int\mathrm{d}^3\bm{r}\,\rho(\bm{r})\int\limits_0^1\mathrm{d}\eta\,\big(\bm{r}\cdot\bm{E}(\bm{R}+\eta\bm{r})\big)+\text{H.c.}\,.
    \label{interaction_Hamiltonian_integral}
\end{equation}

To proceed, we expand the spatial electric field profile $\bm{E}(\bm{R}+\eta\bm{r})$ around the center of mass position $\bm{R}$, provided $r\ll R$. Unlike the previous studies~\cite{Babiker_2002,Alexandrescu_2006,Mondal_2014,Mukherjee_2018} that considered the Cartesian Taylor expansion, we employ the \textit{spherical expansion}~\cite{Weniger_2002,Weniger_2000,Weniger_2005}
\begin{align}
    &\bm{E}(\bm{R}+\eta\bm{r})=\exp(\eta\bm{r}\cdot\bm{\nabla}_{\bm{R}})\bm{E}(\bm{R})\nonumber\\
    &=\sum\limits_{n,l,m}c_{n,l}(\eta r)^{2n+l}Y^{\ast}_{l,m}(\Omega_{\bm{r}})\big[\mathcal{R}_{l,m}(\bm{\nabla}_{\bm{R}}) \bm{E}(\bm{R})\big]\,,
    \label{eq_Taylor}    
\end{align}
where $c_{n,l}=\frac{\pi 2^{l+2}\kappa^{2n}(l+n)!}{n!(2l+2n+1)!}$, $n,l\geq 0$, $|m|\leq l$ and $\kappa$ is the wavenumber of the electric field. $\mathcal{R}_{l,m}(\bm{\nabla}_{\bm{R}})$ are the solid harmonics of the gradient operator $\bm{\nabla}_{\bm{R}}$, also known as spherical tensor gradient operators. Their detailed overview is the subject of Refs.~\cite{Weniger_2000,Weniger_2002,Weniger_2005}. For the sake of simplicity, we refer to the term $[\mathcal{R}_{l,m}(\bm{\nabla}_{\bm{R}}) \bm{E}(\bm{R})]$ as the \textit{spherical gradient}. For differentiable optical fields $\bm{E}(\bm{R})$ and specific values of $l$ and $m$, it can be calculated analytically, after expanding the solid harmonics in Cartesian coordinates
\begin{align}
    &\mathcal{R}_{l,m}(\bm{\nabla}_{\bm{R}})=\sqrt{\frac{(2l+1)(l+m)!(l-m)!}{4\pi}}\nonumber\\
    &\times \sum\limits_{k}\frac{(-\partial_{X}-i\partial_{Y})^{m+k}(\partial_{X}-i\partial_{Y})^{k}\partial_{Z}^{l-m-2k}}{2^{m+2k}(m+k)!k!(l-m-2k)!}\,,
    \label{eq_spherical_gradient}
\end{align}
where $\partial_{X}$, $\partial_{Y}$ and $\partial_{Z}$ are the Cartesian components of the gradient vector $\bm{\nabla}_{\bm{R}}$, $\max(-m,0)\leq k\leq\lfloor\frac{l-m}{2}\rfloor$, with $\lfloor x \rfloor$ being the floor function yielding the largest integer less than $x$. 

We substitute the electric field expansion~\eqref{eq_Taylor} into the Hamiltonian~\eqref{interaction_Hamiltonian_integral}. We consider the rotating-frame charge distribution, given by  Eq.~\eqref{eq_charge_density_mol}, and apply the rotation rule~\eqref{eq_sph_harm_rotation} to obtain the distribution~$\rho(\bm{r},\hat{\Omega})$ in the laboratory coordinates. After integrating over $\eta$ and $\bm{r}$, and taking the limit $\chi\to 0$, we obtain the interaction Hamiltonian~(for details refer to App.~\ref{app_hamiltonian})
\begin{widetext}
\begin{equation}
    \mathcal{H}_{\text{int}}=\sum\limits_{l,m,\sigma,\mu}\gamma_{l,m,\sigma} \hat{\alpha}_{l+1,\mu} \big[\mathcal{R}_{l,m}(\bm{\nabla}_{\bm{R}}) E_{\sigma}(\bm{R})\big]\Big|_{\bm{R}=\hat{\bm{R}}}%
    \begin{cases}
    D^{l+1}_{m-\sigma,\mu}\big(\hat{\Omega}\big)-(-1)^{\mu} D^{l+1}_{m-\sigma,-\mu}\big(\hat{\Omega}\big)&\mu<0\\
    \sqrt{2} D^{l+1}_{m -\sigma,0}(\hat{\Omega})&\mu=0\\
    D^{l+1}_{m -\sigma,-\mu}\big(\hat{\Omega}\big)+(-1)^{-\mu} D^{l+1}_{m - \sigma,\mu}\big(\hat{\Omega}\big)& \mu>0\end{cases}+\text{H.c.}\,,
\label{eq_interaction_Hamiltonian_final}
\end{equation}
\end{widetext}
where $\gamma_{l,m,\sigma} = c_{0,l}\,C^{l,m}_{l+1,m - \sigma;1,\sigma}/\sqrt{4l^2+6l+1}$, with $C^{L,M}_{l,m;l',m'}$ are the Clebsch-Gordan coefficients~\cite{Varshalovich}, and $\{E_{\pm}(\bm{R}),E_0(\bm{R})\}\equiv\{(E_{x}(\bm{R})\pm iE_{y}(\bm{R}))/\sqrt{2},E_{z}(\bm{R})\}$ are the circular polarization components of the electric field $\bm{E}(\bm{R})$. Operators $\hat{\bm{R}}$, $\hat{\alpha}_{l+1,\mu}$ and $\hat{\Omega}$ act respectively on the center-of-mass, vibrational and rotational states of the particle.

The Hamiltonian~\eqref{eq_interaction_Hamiltonian_final} is the major result of our model. Provided all of its building blocks can be properly defined, it is valid for \textit{any rotating particle} and \textit{any profile of the optical field}. The main idea behind $\mathcal{H}_{\text{int}}$ is to absorb most of the light-matter interaction's complexity into the definitions of multipole moments $\hat{\alpha}_{l+1,\mu}$ and electric field components $E_\sigma(\bm{R})$. Afterwards, the resulting expression~\eqref{eq_interaction_Hamiltonian_final} by its design reveals the angular momentum exchange and the corresponding selection rules. Notably, the Hamiltonian~\eqref{eq_interaction_Hamiltonian_final} describes only the coupling of light to rotational and vibrational degrees of freedom that are decoupled from one another. In the case, when the rotations and vibrations of the particle are coupled, one needs to consider the ro-vibrational Hamiltonian $\mathcal{H}_\text{RV}$ as an additional perturbation to the aforementioned Hamiltonian $\mathcal{H}$, which would intermix the excitations, induced by $\mathcal{H}_\text{int}$.

Motivated by ro-vibrational spectroscopy, we introduce two assumptions that are characteristic of the optical fields used in such experiments and substantially simplify the derivations. First of all, we restrict ourselves to the electric field $\bm{E}(\bm{R})$ of a \textit{non-tightly focused beam}. It allows us to use the small-angle approximation, i.e., to separate the polarization vector and spatial profile of the electric field: $E_\sigma(\bm{R})=\epsilon_\sigma E(\bm{R})$. The circular polarization~$\sigma$ and spatial profile $E(\bm{R})$ are associated respectively with the spin and OAM of light~\cite{Bliokh_OAM,Andrews_Book}. In the case of a tightly focused beam, where the focal spot spans just a few wavelengths, these two angular momenta are strongly coupled~\cite{Zhao_2007,Bliokh_SOC}, which noticeably complicates the analysis of the AM transfer from the electric field to the particle. Besides that, the selection rules for tightly focused beams depend on the particularities of the electric field and can not be generalized. In particular, such a calculation requires obtaining focal fields that depend on non-analytic integrals of the moments of the apodization function~\cite{Novotny_Hecht}. Nevertheless, tight focusing was shown to improve the OAM transfer, as discussed in Sec.~\ref{sec_conclusion}. This renders the characterization of the interaction of such light sources with matter an important future extension to our theory.

As the second simplification, we assume that the optical field has a characteristic axis. For a beam, one can consider the time average of the Poynting vector and, for a standing wave in a cavity, the axis would be one of the cavity's symmetry axes. We assume that the characteristic field axis is oriented along the laboratory frame $z$-axis. By choosing the specific coordinate system with respect to the electric field orientation, we explicitly define quantization axes for the AM operators, such as $\hat{L}_z$. Therefore, the particular selection rules that we derive, are only valid in the chosen coordinates. The selection rules for a different orientation of coordinate axes can be obtained from our results using the rotational transformation~\eqref{eq_sph_harm_rotation}. In addition, we formulate the selection rules in terms of \textit{magnetic quantum numbers}, i.e., eigenvalues of the $\hat{L}_z$ operator. Since beams and cavity fields are often cylindrically rather than spherically symmetric, they do not have a well-defined value of the azimuthal quantum number ($\hat{\bm{L}}^2$ eigenvalue).

To derive the selection rules, we consider cylindrical coordinates ${\bm{R}=\{R,\Phi,Z\}}$ and expand the electric field profile into Fourier series with respect to the polar angle~$\Phi$: $E(\bm{R})=\sum_M E_M(R,Z)e^{iM\Phi}$. Each term in the series is an eigenvalue of the $\hat{L}_{z,\Phi}=-i\hbar\,\partial/\partial \Phi$ operator with the corresponding magnetic quantum number $M$. The spherical gradient can be straightforwardly calculated, using Eq.~\eqref{eq_spherical_gradient}
\begin{equation}
	\big[\mathcal{R}_{l,m}(\bm{\nabla}_{\bm{R}}) E(\bm{R})\big]=\sum\limits_{M}\tilde{E}_{M,l}(R,Z)e^{i(M+m)\Phi}\,.
	\label{eq_sph_grad_fourier}
\end{equation}
Applying the spherical gradient operator changes the Fourier amplitudes: $E_M(R,Z)\to\tilde{E}_{M,l}(R,Z)$, and shifts the magnetic quantum numbers: $M\to M+m$.

\begin{figure}[ht]
	\centering
    \vspace{0.25\baselineskip}
	\includegraphics[width=\linewidth]{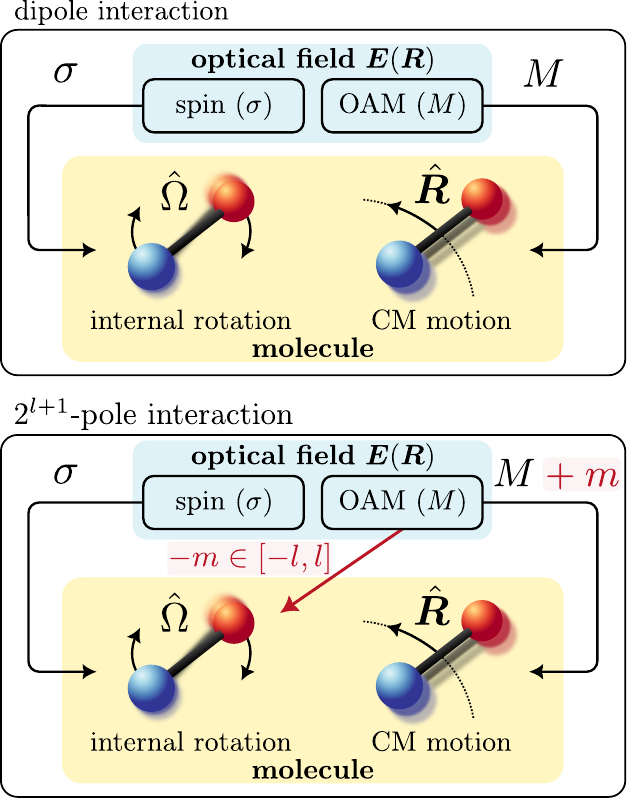}
    \caption{Summary of the angular momentum exchange rules revealed by the Hamiltonian~\eqref{eq_interaction_Hamiltonian_final}. This general pattern applies to any beam with an insignificant spin-orbit coupling of light. (top) Within the electric-dipole interaction, only the spin $\sigma$ of the optical field $\bm{E}(\bm{R})$ can be transferred to the molecular (internal) rotation, described by the angle $\hat{\Omega}$. OAM of the beam, if present, is fully absorbed by the center-of-mass motion $(\hat{\bm{R}})$. (bottom) Within higher orders of the light-matter interaction, OAM of the electric field can be transferred to the molecular rotation. Within the $2^{l+1}$-pole interaction, this transfer is limited to $l$ quanta at most and the transfer probability is determined by the spherical gradient~\eqref{eq_spherical_gradient} of the electric field.}
	\label{fig_interaction_scheme}
\end{figure}

The non-zero spherical gradient~\eqref{eq_sph_grad_fourier} for certain $l$ and $m$ represents the transfer of $M+m$ quanta of angular momentum from the beam to the center-of-mass motion $(\hat{\bm{R}})$. The ability to directly associate a derivative from the Taylor series to the angular momentum transfer is the main benefit of using the spherical expansion in Eq.~\eqref{eq_Taylor} instead of the Cartesian one. Similarly, Wigner D-matrices $D^{l+1}_{m-\sigma,\pm\mu}\big(\hat{\Omega}_{\text{mol}}\big)$ describe the transfer of $\sigma-m$ quanta of AM from the electric field to the molecular (internal) rotation~$(\hat{\Omega}_{\text{mol}})$. Note, that for the given value of $l$, the angular momentum transfer to the internal rotation is bounded: $|\sigma-m|\leq l+1$. This happens, because spherical harmonics, used in the expansion~\eqref{eq_charge_density_mol}, are associated with the angular momentum $\lambda=l+1$, and this property is inherited by the spherical multipoles $\hat{\alpha}_{l+1,\mu}$. The aforementioned selection rules, summarized in Fig.~\ref{fig_interaction_scheme}, satisfy the \textit{angular momentum conservation}: $M+\sigma$ quanta of AM of the electric field get redistributed into $M+m$ quanta of center-of-mass AM and $\sigma-m$ quanta of molecular AM.  Apart from the angular momentum exchange, Eq.~\eqref{eq_interaction_Hamiltonian_final} infers two major corollaries.

%%%%%%%%%%%%%%%%%%%%%%%%%%%%%%%%%%%%%%%%%%%%%%%%%%
%%%%%%%%%%%%%%%%%%%%%%%%%%%%%%%%%%%%%%%%%%%%%%%%%%

If one considers only the electric-dipole contribution to the light-matter interaction, the remaining terms in the Hamiltonian~\eqref{eq_interaction_Hamiltonian_final} are those with $l=0$. The spherical gradient $[\mathcal{R}_{0,0}(\bm{\nabla}_{\bm{R}}) E(\bm{R})]|_{\bm{R}=\hat{\bm{R}}}\propto E(\hat{\bm{R}})$ is independent of summation indices and can be extracted outside the sum. As a result, the electric field profile $E(\hat{\bm{R}})$ is unable to influence molecular rotational transitions. In this case, molecular transitions are affected only by the circular polarization $\sigma$, i.e., spin of light. This result is in agreement with the previous studies~\cite{Babiker_2002,Mondal_2014}. In fact, for a diatomic molecule, described by the charge density~\eqref{eq_charge_density_diatomic}, the contribution to the light-matter interaction from the electric dipole reads
\begin{equation}
    \mathcal{H}_{\text{int}}^{\text{dip}}=-\sqrt{\frac{4\pi}{3}}\hat{d}^{\text{vib}}_{z}E(\hat{\bm{R}})\sum\limits_{\sigma}\epsilon_{\sigma}Y_{1,\sigma}(\hat{\Omega})+\text{H.c.}\,,
    \label{eq_Hamiltonian_dip}
\end{equation}
which is identical to the result of Ref.~\cite{Mondal_2014}. In the case of a tightly focused beam, the polarization-dependent electric field cannot be extracted outside the sum in Eq.~\eqref{eq_Hamiltonian_dip}. As a result, apart from the spin, the spatial dependence of the polarization, which is characteristic of the spin and OAM interaction, also affects the molecular (internal) rotation at the dipole level of interaction.

In contrast to the dipole case, if one considers the electric-quadrupole interaction,~i.e.,~terms with $l=1$ in Eq.~\eqref{eq_interaction_Hamiltonian_final}, the spherical gradient $[\mathcal{R}_{1,m}(\bm{\nabla}_{\bm{R}})E(\bm{R})]|_{\bm{R}=\hat{\bm{R}}}$ is coupled to the D-matrix of the molecular orientation $\hat{\Omega}$ through the summation index $m$. In this case, apart from the polarization $\sigma$, molecular rotational transitions are also determined by the gradients of the electric field, which are thoroughly discussed in Sec.~\ref{sec_LG_beams}. For a diatomic molecule, the electric-quadrupole interaction term is given by
\begin{align}
    \mathcal{H}_{\text{int}}^{\text{quad}}&=\pi\sqrt{2/3}\,\hat{Q}^{\text{vib}}_{z,z}\sum\limits_{m\sigma}\epsilon_{\sigma}C^{1,m}_{2,m-\sigma;1,\sigma}Y^{\ast}_{2,m-\sigma}(\hat{\Omega})\nonumber\\
    &\times\big[\mathcal{R}_{1,m}(\bm{\nabla}_{\bm{R}}) E(\bm{R})\big]\Big|_{\bm{R}=\hat{\bm{R}}}+\text{H.c.}\,.
    \label{eq_Hamiltonian_quad}
\end{align}
This clearly illustrates that within the electric-quadrupole (as well as the higher-order) interaction, the spatial profile of the beam, hence its OAM, affects the selection rules on molecular rotational transitions.

\section{Laguerre-Gaussian beams and the enhancement of rotational transitions}
\label{sec_LG_beams}

To analyze in detail, how the spatial profile of the electric field $E(\bm{R})$ affects molecular rotational transitions, one needs to choose the specific beam profile. We consider an electric field that can be expressed as a superposition of Laguerre-Gaussian (LG) modes. These modes are solutions to the paraxial Helmholtz equation in cylindrical coordinates $\bm{R}=\{R,\Phi,Z\}$. The magnitude of the electric field in a LG mode is given by~\cite{Allen_1992}
\begin{align}
    &E_{P,M}(\bm{R})=\gamma_{P,M}\bigg(\frac{R}{\omega_Z}\bigg)^{|M|}\exp[-\frac{R^2}{\omega^2_Z}]e^{-iM\Phi}e^{-i\kappa Z}\nonumber\\
    &\times\Bigg\{\frac{\omega_0}{\omega_Z}\exp[-i\bigg(\frac{\kappa R^2}{2\mathbb{R}_Z}-\psi_Z\bigg)]\mathcal{L}^{|M|}_{P}\bigg(\frac{2R^2}{\omega^2_Z}\bigg)\Bigg\}\,,
    \label{eq_EF_Allen}
\end{align}
where $\kappa=2\pi n/\lambda_{\text{beam}}$ is the wavenumber, $n$ is the refraction index, $\lambda_{\text{beam}}$ is the wavelength of light, and $\mathcal{L}^{j}_{i}(x)$ with $i=0,1,2,...$ and $j\in\mathbb{R}$ are the generalized Laguerre polynomials. The beam is focused at $Z=0$, with the waist function $\omega_Z=\omega_0\sqrt{1+(Z/Z_R)^2}$ along the $z$-axis, where $\omega_0$ is the waist at the focus and ${Z_R=\pi\omega_0^2 n/\lambda_{\text{beam}}}$ is the Rayleigh length. The radius of the wavefront curvature $\mathbb{R}_Z=Z(1+(Z_R/Z)^2)$ and the Gouy phase $\psi_Z=(2P+|M|+1)\atan(Z/Z_R)$. The electric field~\eqref{eq_EF_Allen} is normalized in the sense of the Dirac delta-function with respect to the axial coordinate $Z$, and to unity with respect to coordinates $R$ and $\Phi$, with the normalization constant $\gamma_{P,M}=\sqrt{2^{|M|+1}P!/(\pi(P+|M|)!)}$.

According to Ref.~\cite{Allen_1992}, the angular momentum density of the electric field~\eqref{eq_EF_Allen} precesses about the beam axis. Its projection on the $z-$axis is well defined and depends only on the beam parameters, but its projection on the transverse plane changes its amplitude and orientation depending on coordinates $R$ and $\Phi$. As a consequence, the total angular momentum of the beam, in a sense of the $\hat{\bm{L}}^2$ operator eigenvalue cannot be defined. Instead, the electric field $E_{P,M}(\bm{R})$ is an eigenfunction of the $\hat{L}_{z,\Phi}$ operator, defined in Sec.~\ref{sec_hamiltonian}, with the corresponding magnetic quantum number $M$. Eq.~\eqref{eq_EF_Allen} with $M=0$ describes a Gaussian beam without OAM, and serves within this manuscript as a benchmark representation of the non-helical light.

As discussed in Sec.~\ref{sec_hamiltonian}, the leading term in the Hamiltonian~\eqref{eq_interaction_Hamiltonian_final}, which describes the \textit{OAM transfer to the molecular (internal) rotation}, is the electric-quadrupole interaction,~i.e.,~the general case of Eq.~\eqref{eq_Hamiltonian_quad}. In what follows, we choose the absolute value of the spherical gradient $\big|\mathcal{R}_{1,m}(\bm{\nabla}_{\bm{R}})E_{P,M}(\bm{R})\big|$ with $m=\pm1$ as the quantitative measure of this OAM transfer. For simplicity, we abuse the notation $\bm{R}=\hat{\bm{R}}$, while preserving its meaning in the Hamiltonian~\eqref{eq_interaction_Hamiltonian_final}, namely, that $\bm{R}$ is the coordinate of the molecular center of mass. The action of the spherical gradient on the electric field $E_{P,M}(\bm{R})$ is readily given by Eq.~\eqref{eq_sph_grad_fourier}, where the transformed Fourier amplitude can be calculated using Eq.~\eqref{eq_spherical_gradient}. Note that we disregard the gradient with $m=0$, as it describes a different angular momentum transfer process. Namely, the transfer of the entire OAM of the field to the center-of-mass motion, accompanied by the transfer of the spin of light to the molecular rotation (see Fig.~\ref{fig_interaction_scheme}).

\begin{figure}[ht]
	\centering
	\includegraphics[width=\linewidth]{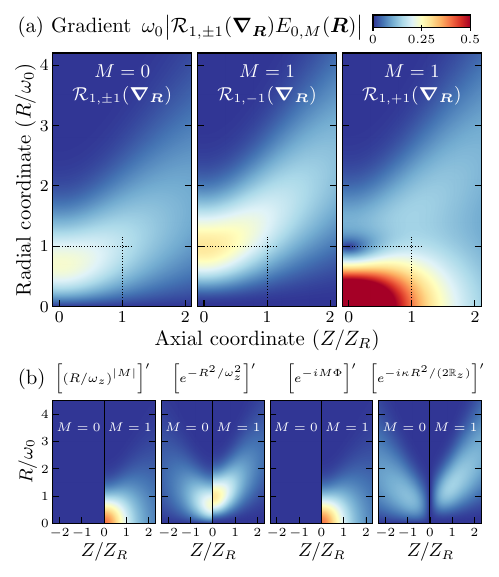}
    \caption{Spherical gradients $[\mathcal{R}_{1,\pm 1}(\bm{\nabla}_{\bm{R}})E(\bm{R})]$ quantify the OAM transfer to the molecular (internal) rotation. \mbox{(a) Dependence of} the spherical gradients on radial $(R)$ and axial $(Z)$ coordinates of the molecular center-of-mass, calculated for the Laguerre-Gaussian beam profile $E_{0,M}(\bm{R})$. For the optical field with non-zero helicity $M=1$, the gradient is substantially enhanced around the central (axial) focal point of the beam $(|R|\leq \omega_0,|Z|\leq Z_R)$, when compared with the non-twisted light $(M=0)$. This enhancement is present, only if $\text{sgn}(m)=\text{sgn}(M)$. (b)~Spherical gradients (denoted here by the $\prime$-symbol) of the four main components of the Laguerre-Gaussian mode~\eqref{eq_EF_Allen}. The left (right) half of each panel corresponds to a beam without (with) the OAM. In all cases the results are equivalent for both $m = \pm 1$ gradients. In contrast to the non-helical light, gradients of the in-center radial distribution (leftmost panel) and spiral beam phase (third panel from the left) have sizeable amplitudes in the case of a vortex beam. Even though $m=\pm 1$ gradients are equivalent for all individual components, their corresponding contributions constructively interfere for $m M > 0$, while they destructively interfere for $m M < 0$ resulting in the overall enhancement and suppression observed in (a).}
	\label{fig_spherical_gradients}
\end{figure}

In Fig.~\ref{fig_spherical_gradients}(a), we plot the dependence of spherical gradients $|\mathcal{R}_{1,\pm1}(\bm{\nabla}_{\bm{R}})E_{0,M}(\bm{R})|$ on (cylindrical) center-of-mass coordinates $R$ and $Z$. For simplicity, we demonstrate our findings for ${P=0}$, however, other values of $P$ display qualitatively similar physics. As expected, the shape of the gradient profile roughly follows the shape of the beam profile~\eqref{eq_EF_Allen}. We observe that, for the optical field with $M=1$, the gradient is substantially enhanced, in comparison with the non-twisted light with $M=0$. The enhancement is present, only if $\text{sgn}(m)=\text{sgn}(M)$. In the opposite case, the OAM transfer is only marginally increased, due to the different normalization constants $\gamma_{P,M}$ and radial distributions for helical and non-helical beams. Notably, the enhancement is localized around the center (axis) and focus of the beam, i.e., in the region with $|R|\leq \omega_0$ and $|Z|\leq Z_R$. This indicates that the experimental observation of such an enhancement would require the molecules to be placed around this region in the beam, as further discussed in Sec.~\ref{sec_experiment}.

%%%%%%%%%%%%%%%%%%%%%%%%%%%%%%%%%%%%%%%%%%%%%%%%%%
%%%%%%%%%%%%%%%%%%%%%%%%%%%%%%%%%%%%%%%%%%%%%%%%%%

To figure out the origin of the observed enhancement, we consider different fundamental components of the electric field~\eqref{eq_EF_Allen} separately. In particular, we distinguish between the spiral beam phase:~$e^{-iM\Phi}$, the in-center radial distribution:~$(R/\omega_Z)^{|M|}$, relevant for ${R\ll\omega_0}$, the off-center exponential decay:~$e^{-R^2/\omega_Z^2}$, relevant for $R\gg\omega_0$, as well as the wavefront curvature: $e^{-i\kappa R^2/(2\mathbb{R}_{Z})}$. The non-zero curvature of the wavefront becomes relevant in the \textit{out-of-focus} regime, i.e.,~when $Z\gg Z_R$. It effectively deforms the radial distribution of the optical field in the transverse plane at a given axial coordinate $Z$, leading to a change in the gradient.

In Fig.~\ref{fig_spherical_gradients}(b), we plot the contributions to the spherical gradient $|\mathcal{R}_{1,\pm 1}(\bm{\nabla}_{\bm{R}})E_{0,M}(\bm{R})|$ for each of the aforementioned four components of the electric field~\eqref{eq_EF_Allen}. For brevity, we denote the spherical gradient by the \mbox{$\prime$-symbol.} As expected, gradients of the exponential decay and wavefront curvature term are not sensitive to the change in beam vorticity $M$ and only change marginally. The in-center radial distribution (power term) and spiral phase, in contrast, are absent for the non-twisted light, but their gradients have sizeable amplitudes in the case of a helical beam. The gradients, presented in Fig.~\ref{fig_spherical_gradients}(b), are identical for both values of the index $m=\pm 1$. However, the way these contributions add up to form the overall spherical gradient $|\mathcal{R}_{1,\pm 1}(\bm{\nabla}_{\bm{R}})E_{0,M}(\bm{R})|$ is different. The relative phase between the terms depends on the index $m$. If $\text{sgn}(m)=\text{sgn}(M)$, all contributions are ``in-phase'', which results in the enhancement of the spherical gradient, depicted in Fig.~\ref{fig_spherical_gradients}(a).

Our findings indicate that the overall magnitude of the OAM transfer is an interplay between spherical gradients of the aforementioned fundamental components of the optical field. To illustrate this fact in a simple way, we calculate the spherical gradient $|\mathcal{R}_{1,\pm1}(\bm{\nabla}_{\bm{R}})E_{0,M}(\bm{R})|$ analytically. The expression for the full electric field profile~\eqref{eq_EF_Allen} is too cumbersome and we omit it here. Instead, we consider an approximation to the electric field~\eqref{eq_EF_Allen}
\begin{equation}
    E^{\text{foc}}_M(\bm{R})=\gamma_{0M}\left( R/\omega_0 \right)^{|M|}e^{-R^2/\omega^2_0}e^{-iM\Phi}e^{-i\kappa Z}\,.
    \label{eq_EF_approximation_1}
\end{equation}
This corresponds to a LG beam with $P=0$ within the \textit{in-focus approximation}, which is valid, when $Z\ll Z_{R}$, thus $\omega_{Z}\to\omega_0$, $1/R_{Z}\to0$ and $\psi_{Z}\to 0$. We choose this approximation, because, as indicated in Fig.~\ref{fig_spherical_gradients}(a), the enhancement of the spherical gradient occurs in the focal region of the beam. Unlike previous studies~\cite{Alexandrescu_2006,Mondal_2014}, approximation~\eqref{eq_EF_approximation_1} includes the exponential decay, as it is needed for the accurate analysis. The comparison between different approximations is given in App.~\ref{app_EF_approximations}. The spherical gradients of $E^{\text{foc}}_M(\bm{R})$ read 
\begin{align}
    &\big[\mathcal{R}_{1,\pm1}(\bm{\nabla}_{\bm{R}}) E^{\text{foc}}_{M}(\bm{R})\big]=\nonumber\\
    &=-\sqrt{\frac{3}{8\pi}}e^{\pm i\Phi}\bigg[\frac{M}{R}\pm\bigg(\frac{|M|}{R}-\frac{2R}{\omega_0^2}\bigg)\bigg]E^{\text{foc}}_{M}(\bm{R})\,.
    \label{eq_gradient}
\end{align}
Looking back at the full electric field profile~\eqref{eq_EF_Allen}, the three terms in the square brackets of Eq.~\eqref{eq_gradient} correspond, in this order, to spherical gradients of the spiral beam phase, in-center radial distribution and off-center exponential decay. The former two contributions are absent in the case of beam with zero vorticity. In the case of a vortex beam, they either interfere constructively (if $m M >0$) or destructively (if $m M < 0$). Notice, that the latter term is present for a Laguerre-Gaussian beam with any value of $M$, implying that, in the electric-quadrupole order of the light-matter interaction, the OAM transfer to the molecular (internal) rotation is happening for both helical and non-helical light. 

Overall, we conclude that the sizeable enhancement of the spherical gradient, which is the quantitative indicator of the OAM transfer to the molecular rotation, is enabled by the components of the optical field that are present only in beams with non-zero helicity.

\section{Ro-vibrational transition rates}
\label{sec_transition_rates}

The enhancement of the OAM transfer from the beam to the molecular (internal) rotation, discussed in Sec.~\ref{sec_LG_beams}, could be experimentally verified using ro-vibrational spectroscopy. Here, we calculate the rates of the corresponding ro-vibrational transitions.

We assume that the state of a single molecule can be described by the following tensor product: ${\vert\Psi\rangle=\vert\psi^{\text{vib}}_\nu\rangle\vert\psi^{\text{rot}}_{J,N_1,N_2}\rangle\vert\psi^{\text{CM}}_{\bm{R}_{\text{CM}}}\rangle}$. This implies that we neglect weak correlation effects, like correlations stemming from the ro-vibrational coupling (see App.~\ref{app_extensions}). We leave the vibrational state $\ket{\psi^{\text{vib}}_\nu}$ implicit, since we calculate the transition multipole moments directly using numerical methods, as discussed in Sec.~\ref{sec_multipole_moments}. The rotational state $\ket{\psi^{\text{rot}}_{J,N_1,N_2}}=\ket{J,N_1,N_2}$ is a generic state in the angular momentum basis. In particular, it is the simultaneous eigenstate of the laboratory-frame $\hat{\bm{J}}^2$ and $\hat{J}_z$, and molecular-frame~$\hat{J}^{\prime}_z$ angular momenta of the molecule. The corresponding wave function in the basis of molecular angle states $\ket{\Omega_\text{mol}}$, i.e., eigenstates of the molecular orientation operator $\hat{\Omega}_{\text{mol}}$, reads 
\begin{equation}
    \bra{\Omega_{\text{mol}}}\ket{\psi^{\text{rot}}_{J,N_1,N_2}}=D^J_{N_1,N_2}(\Omega_{\text{mol}})\,,
\end{equation}
where $D^J_{N_1,N_2}(\Omega_\text{mol})$ are the irreducible representations of the molecular rotation group (see App.~\ref{app_molecular_rotation_group} for details). Provided that the definition of $\Omega_\text{mol}$ already embeds the molecular (symmetry) point group $\mathcal{G}_\text{mol}$, one can simply use the Wigner D-matrices as representations.

For most molecules at relevant experimental temperatures, the thermal de-Broglie wavelength is significantly smaller than the wavelength of the probe light. For instance, for a CS molecule at $T=20\,\text{K}$: the de-Broglie wavelength is $\lambda_{\text{dB}}=h\sqrt{\beta/(2\pi \mathbb{M})}\approx 0.6\,\text{\AA}$, where $\mathbb{M}$ is the molecular mass,  $\beta=1/(k_{\text{B}}T)$ is the Boltzmann factor with the Boltzmann constant $k_{\text{B}}$. At the same time, the wavelength corresponding to the lowest vibrational transition is $\lambda_{\text{beam}}\sim8\,\text{µm}$. For a non-tightly focused beam, like the LG beam~\eqref{eq_EF_Allen}, the beam waist is larger than the wavelength, leading to $\omega_0>\lambda_{\text{beam}}\gg\lambda_{\text{dB}}$. Therefore, the wave function that describes the center-of-mass position inside the beam can be approximated by the three-dimensional $\delta$-function
\begin{equation}
    \bra{\bm{R}}\ket{\psi^{\text{CM}}_{\bm{R}_{\text{CM}}}}=\delta^{(3)}(\bm{R}-\bm{R}_{\text{CM}})\,,
\end{equation}
where $\ket{\bm{R}}$ is an eigenstate of the center-of-mass position operator $\hat{\bm{R}}$.

The amplitude of the transition between the initial $\vert\Psi_i(\bm{R}^i_\text{CM})\rangle$ and final $\vert\Psi_f(\bm{R}^f_\text{CM})\rangle$ molecular states with respect to the full light-matter interaction Hamiltonian~\eqref{eq_interaction_Hamiltonian_final} reads
\begin{align}
    &\mathcal{M}_{i\to f}(\bm{R}^i_\text{CM},\bm{R}^f_\text{CM})=\langle\Psi_f(\bm{R}^f_\text{CM})\vert\mathcal{H}_{\text{int}}\vert\Psi_i(\bm{R}^i_\text{CM})\rangle\nonumber\\
    &=\sum\limits_{nlm\mu}\mathcal{I}^{\text{vib};\,n,l,\mu}_{\nu,\nu'}\,\mathcal{I}^{\text{CM};\,l,m}_{\bm{R}^i_{\text{CM}},\bm{R}^f_{\text{CM}}}\,\mathcal{I}^{\text{rot};\,n,l,m,\mu}_{J,N_1,N_2,J',N'_1,N'_2}\,,
    \label{eq_transition_matrix}
\end{align}
where states $\lvert\Psi_{i(f)}(\bm{R}^{i(f)}_\text{CM})\rangle$ are characterised by sets of quantum numbers: $i(f)\equiv\{\nu^{(\prime)},J^{(\prime)},N_1^{(\prime)},N_2^{(\prime)}\}$, and the vibrational integral $\mathcal{I}^{\text{vib};\,n,l,\mu}_{\nu,\nu'}=\bra{\psi^{\text{vib}}_{\nu'}}\alpha_{l+1,\mu}\ket{\psi^{\text{vib}}_\nu}$ can be expressed through the transition multipole moments, as discussed in Sec.~\ref{sec_multipole_moments}. The rotational and center-of-mass integrals are
\begin{widetext}
\begin{equation}
    \mathcal{I}^{\text{rot};\,n,l,m,\mu}_{J,N_1,N_2,J',N'_1,N'_2}=\bigg[\sum\limits_{\sigma}\frac{8\pi^2}{2l+3}\gamma_{n,l,m,\sigma}\epsilon_{\sigma}C^{J',N_1'}_{l+1,m-\sigma;J,N_1}\bigg]\begin{cases}
    C^{J',N_2'}_{l+1,\mu;J,N_2}-(-1)^\mu C^{J',N_2'}_{l+1,-\mu;J,N_2}&\mu<0\\
    \sqrt{2} C^{J',N_2'}_{l+1,0;J,N_2}&\mu=0\\
    C^{J',N_2'}_{l+1,-\mu;J,N_2}+(-1)^{-\mu} C^{J',N_2'}_{l+1,\mu;J,N_2}& \mu>0\end{cases}\,,
\end{equation}
\begin{equation}
    \mathcal{I}^{\text{CM};\,l,m}_{\bm{R}^i_{\text{CM}},\bm{R}^f_{\text{CM}}}=\bra{\psi^{\text{CM}}_{\bm{R}^f_{\text{CM}}}}\big[\mathcal{R}_{l,m}(\bm{\nabla}_{\bm{R}}) E(\bm{R})\big]\Big|_{\bm{R}=\hat{\bm{R}}}\ket{\psi^{\text{CM}}_{\bm{R}^i_{\text{CM}}}}
    =\delta^{(3)}(\bm{R}^i_{\text{CM}}-\bm{R}^f_{\text{CM}})\big[\mathcal{R}_{l,m}(\bm{\nabla}_{\bm{R}^i_{\text{CM}}}) E(\bm{R}^i_{\text{CM}})\big]\,.
\end{equation}
\end{widetext}

After defining the amplitude $\mathcal{M}_{i\to f}(\bm{R}^i_\text{CM},\bm{R}^f_\text{CM})$, we obtain the corresponding transition rate using Fermi's golden rule
\begin{align}
    \Gamma_{i\to f}&(\omega,\bm{R}^i_\text{CM},\bm{R}^f_\text{CM})=\nonumber\\
    &\frac{2\pi}{\hbar}\big|\mathcal{M}_{i\to f}(\bm{R}^i_\text{CM},\bm{R}^f_\text{CM})\big|^2\delta_{E}(\hbar\omega-\Delta E_{i\to f})\,,
    \label{eq_Fermi_rule}
\end{align}
where $\omega$ is the angular frequency of the photon that drives the transition and $\Delta E_{i\to f}=E_f-E_i$ is the energy difference between the initial and final states. The $\delta$-function in Eq.~\eqref{eq_Fermi_rule} denotes the density of energy states and has therefore the units of inverse energy.

\section{Possible experimental scheme}
\label{sec_experiment}

After defining the transition rates of molecular ro-vibrational transitions, we proceed with a suggestion for the generic proof-of-principle experimental scheme. The proposed setup may be capable of revealing the enhancement of ro-vibrational transition amplitudes, induced by the non-zero vorticity of the probe light.

\begin{figure}[p]
	\centering
	\includegraphics[width=\linewidth]{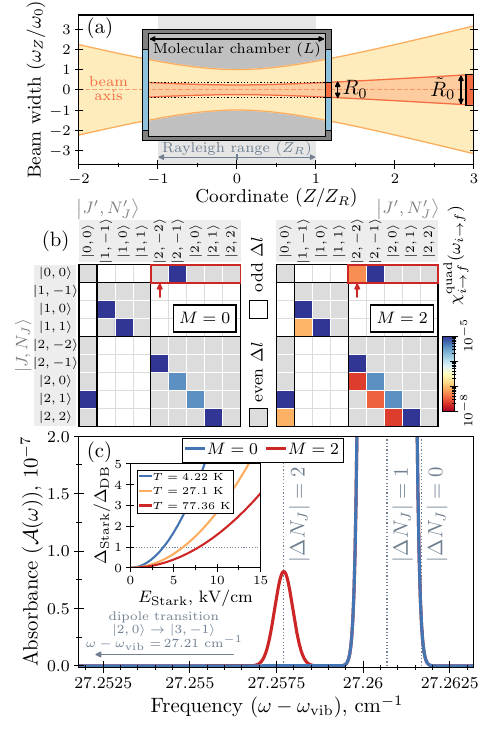}
	\caption{(a)~Schematic of the proof-of-principle experiment to reveal OAM-enabled selection rules. A beam with the focal waist length $\omega_0$ passes through the molecular chamber of length~$L$. The transmitted light is collected by the aperture $\tilde{R}_0$ outside the chamber. If $L\sim Z_R$, most of the molecules that interact with the light which is collected by the aperture, are found within the cylinder, defined by the effective aperture $R_0$~(dotted line). (b)~Matrices of the single-channel attenuation $\chi^{\text{quad}}_{i\to f}(\omega_{i\to f})$ within the electric-quadrupole order of the interaction. The rotational transitions $\ket{J,N_J}\to\ket{J',N_J'}$ are associated with the fundamental vibrational transition $(\nu=0\to\nu'=1)$ of the CS molecule. Compared to a non-helical light~(left), a vortex beam~(right) enables additional transitions with $\chi^{\text{quad}}_{i\to f}(\omega)>10^{-8}$, like those within the rotational O-branch~($\Delta J=2$, boxed).
    (c)~Absorbance spectrum $\mathcal{A}(\omega)$ of CS molecules for parameters: $T=20\,$K, $p=1\,$mbar, $L=15\,$cm, $R_0=150\,\text{µm}$ and $\omega_0=400\,\text{µm}$. The displayed frequency range corresponds to electric-quadrupole transitions within the O-branch of the fundamental vibrational transition~$\omega_{\text{vib}}\approx1250\,\text{cm}^{-1}$. OAM of light induces the substantial enhancement of $|\Delta N_J|=2$ transition. The closest electric-dipole transition~(arrow) is far outside the given frequency range and does not affect the signal. Transitions with different $|\Delta N_J|$ values are separated as a result of applying a static field $E_{\text{Stark}}=7\,\text{kV}/\text{cm}$ along the z-axis. The ratio between the Stark-splitting of $|\Delta N_J|=1$ and $|\Delta N_J|=2$ lines and the characteristic Doppler broadening for different temperatures and amplitudes of the Stark field is presented in the inset.}
	\label{fig_experiment}
\end{figure}

We consider a vacuum chamber of characteristic length $L$ (along the $z$-axis) that contains molecules in the gas phase under pressure $p$ and at temperature $T$. As discussed in Sec.~\ref{sec_LG_beams}, the OAM-induced rotational enhancement depends on the position of the molecular center of mass with respect to the beam axis. For this reason, we suggest measuring the absorbance of light in a spatially-resolved manner. In particular, we propose measuring the ratio between the absorbed and incident power for the molecules, which reside within the optical path of light that is collected by an \textit{adjustable} circular aperture~$\tilde{R}_0$, placed outside the chamber and centered on the beam axis. For small chambers $(L<Z_R)$, such molecules are approximately confined to the cylinder, described by the effective aperture $R_0$~(see Fig.~\ref{fig_experiment}(a)). Note that due to the small average thermal center-of-mass velocity of molecules: $\expval{v_T}\approx\sqrt{3k_\text{B}T/\mathbb{M}}$, their motion  within the chamber is negligible on the time-scale of the photon propagation $(\approx L/c)$. For instance, $\expval{v_T}L/c \approx 44~{\rm nm}$ for $L = 15~{\rm cm}$ and $T = 20~{\rm K}$. 

The experimentally observable absorbance $\mathcal{A}(\omega)$ of photons with angular frequency $\omega$ can be calculated using Beer-Lambert's law~\cite{Tokmakoff_2014}: $\mathcal{A}(\omega)=1-\exp(-\chi_{\text{tot}}(\omega))$, where the total attenuation $\chi_{\text{tot}}(\omega)=\sum_{i,f}\chi_{i\to f}(\omega)$ and $\chi_{i\to f}(\omega)$ is the attenuation, associated with the single ro-vibrational transition $i\to f$, herewith referred to as a channel. Note that here, as in Sec.~\ref{sec_transition_rates}, indices $i$ and $f$ denote only the vibrational and rotational quantum numbers. The single-channel attenuation is defined as the integral: $\chi_{i\to f}(\omega)=\int_{-L/2}^{L/2}\tilde{\chi}_{i\to f}(\omega,Z)\,\mathrm{d}Z$. The local attenuation at the position $Z$ is the ratio: $\tilde{\chi}_{i\to f}(\omega,Z)=\mathbb{P}^{i\to f}_{\text{abs}}(Z)/\mathbb{I}_{\text{inc}}(Z)$, where $\mathbb{I}_{\text{inc}}(Z)$ is the energy transfer rate of the incident electric field, given by
\begin{equation}
    \mathbb{I}_{\text{inc}}(Z)=\frac{c\varepsilon_0}{2}\int_{S_\text{A}}|E(\bm{R})|^2 \,\mathrm{d}^2\bm{R}\,,
\end{equation}
where $\int_{S_\text{A}} \mathrm{d}^2 \bm{R} \equiv \int_0^{R_0} R\,\mathrm{d}R~ \int_{0}^{2\pi} \mathrm{d} \Phi$ and $\mathbb{P}^{i\to f}_{\text{abs}}(Z)$ is the energy flux,~i.e.,~energy transfer rate per unit length, absorbed by molecules, undergoing the transition $i\to f$ and residing in a thin section of the chamber $(\mathrm{d}Z)$. The absorbed energy flux is given by
\begin{align}
    &\mathbb{P}^{i\to f}_{\text{abs}}(Z^i_\text{CM} = Z)=\rho_0\frac{\rho_{\text{B}}(E_i,T)}{\mathcal{Z}}\nonumber\\
    &\times\int_{S_\text{A}}\mathrm{d}^2\bm{R}^{i}_{\text{CM}}\int_{\mathbb{R}^3}\mathrm{d}^3\bm{R}^{f}_{\text{CM}}\Big(\hbar\omega\,\Gamma_{i\to f}(\omega,\bm{R}^i_\text{CM},\bm{R}^f_\text{CM})\Big)\,,
    \label{eq_absorbed_flux}
\end{align}
where $\rho_0=\beta p$ is the equilibrium density of molecules, $\rho_{\text{B}}(E_i,T)=\exp(-\beta E_i)$ is the Boltzmann probability to occupy the initial state, ${\mathcal{Z}=\sum_{i}\rho_{\text{B}}(E_i,T)}$ is the canonical partition function. The integral over the initial center-of-mass position ($\bm{R}^i_\text{CM}$) is two-dimensional and covers the effective aperture $R_0$~(see Fig.~\ref{fig_experiment}(a)). The integral over the final position ($\bm{R}^f_\text{CM}$) is three-dimensional and covers the whole space. At a finite temperature $T$, spectral lines are broadened, so instead of a $\delta$-function, used in the definition of the transition rate~\eqref{eq_Fermi_rule}, we employ the Doppler-broadened line profile
\begin{align}
    &\delta_E(\hbar\omega-\Delta E_{i\to f}))\to\rho_{\text{DB}}(\omega,\Delta E_{i\to f},T)\nonumber\\
    &=\sqrt{\frac{\beta \mathbb{M}c^2}{2\pi (\Delta E_{i\to f})^2}}\exp(-\frac{\beta \mathbb{M}c^2(\hbar\omega-\Delta E_{i\to f})^2}{2(\Delta E_{i\to f})^2})\,.
\end{align}

To demonstrate our findings, we consider a gas of CS molecules with mass $\mathbb{M}=44.076\,\text{amu}$ at pressure $p=1\,\text{mbar}$ and temperature $T=20\,\text{K}$. The rotational state of a linear molecule, like CS, can be described by two quantum numbers $\{J,N_J\}$ instead of three $\{J,N_1,N_2\}$, as required in the case of an asymmetric molecule~(see App.~\ref{app_molecular_rotation_group}). For the molecular state, defined in Sec.~\ref{sec_transition_rates}, the difference in energy between the initial and final states $\Delta E_{i\to f}$ is the difference in rotational and vibrational energies. The combined rotational-vibrational energy for a linear molecule can be calculated using the Dunham expansion~\cite{Dunham_1932} with coefficients for CS provided in Ref.~\cite{Ram_1995}. We consider irradiating molecules with a Laguerre-Gaussian beam~\eqref{eq_EF_Allen} with a width at focus $\omega_0=400\,\text{µm}$. The beam is circularly polarized with $\epsilon_{-1}\neq 0$, i.e., $\sigma=-1$ in Eq.~\eqref{eq_interaction_Hamiltonian_final}. The wavelength of light is close to the fundamental vibrational transition of CS,~i.e.,~the beam excites the transition $\nu=0\to\nu'=1$ with the characteristic angular frequency $\omega_{\text{vib}}=1250\,\text{cm}^{-1}$. Note that, compared to Eq.~\eqref{eq_EF_Allen}, the Rayleigh range in the experimental setup is given by: $Z_{R}=\pi \omega_0^2/(\lambda \mathfrak{M}^2)$, where $\mathfrak{M}^2$ is the beam quality factor~\cite{Siegman_1998}. For a Laguerre-Gaussian beam: $\mathfrak{M}^2\sim2|M|+1$. We calculate the corresponding (vibrational) transition dipole $|d^{0,1}_z|=0.156\,\,\text{D}$ and quadrupole $|Q^{0,1}_{zz}|=0.234\,\,\text{D}\cdot\text{\AA}$ moments using density functional theory, as described in Sec.~\ref{sec_multipole_moments}. In an actual experiment, the transition multipole moments can also be measured directly. Our numerical analysis has shown that the chamber length $L=15\,\text{cm}$ and the effective aperture $R_0=150\,\text{µm}$ maximize the overall transition amplitudes. 

We focus on the electric-quadrupole order of the light-matter interaction, described by transition amplitudes~\eqref{eq_transition_matrix} with $l=1$. We plot single-channel attenuation matrices $\chi^\text{quad}_{i\to f}(\omega_{i\to f})$, where $\hbar\omega_{i\to f}=\Delta E_{i\to f}$, for LG beams with $M=0$ and $M=2$ in Fig.~\ref{fig_experiment}(b). To visually distinguish ``forbidden'' and ``allowed'' transitions, we set the cutoff for the attenuation to $10^{-8}$. Despite the fact, that matrices for helical and non-helical light have very similar structure, our calculation reveals additional ``forbidden'' rotational transitions, enabled by the OAM of light. These transitions are the manifestation of the main argument of Sec.~\ref{sec_LG_beams}, namely, that the non-zero helicity of light can substantially enhance the OAM transfer to the molecular (internal) rotation. From Fig.~\ref{fig_experiment}(b), one can notice that the selection rules on the azimuthal quantum number~$J$ remain the same for $M=0$ and $M=2$. In other words, $\Delta J$ transitions are not modified by the OAM of light. As discussed in Sec.~\ref{sec_LG_beams}, the electric field in a vortex beam is an eigenstate of the $\hat{L}_{z,\Phi}$ operator, as it is \textit{cylindrically} symmetric. Quantum number $J$ is the eigenvalue of the $\hat{J}^2$ operator and describes the \textit{spherical} symmetry of the molecular state. Since for the molecule $[\hat{J}_z,\hat{J}^2] = 0$, the beam, which induces an effective $\hat{J}_z-$interaction onto the molecule, is incapable of modifying the spherical symmetry of molecular states.

%%%%%%%%%%%%%%%%%%%%%%%%%%%%%%%%%%%%%%%%%%%%%%%%%%
%%%%%%%%%%%%%%%%%%%%%%%%%%%%%%%%%%%%%%%%%%%%%%%%%%

The inability of a vortex beam to affect the selection rules on the azimuthal quantum number $J$ necessitates an additional requirement for the experiment. In particular, in the aforementioned setup molecular ro-vibrational energies are degenerate with respect to the magnetic quantum number $N_J$. This implies that, by the means of the absorption spectroscopy, one can only measure a weighted sum of single-channel attenuations $\chi^\text{quad}_{i\to f}(\omega_{i\to f})$ with the same frequencies $\omega_{i\to f}$, i.e., the superposition of peaks for all possible $\Delta N_J\in[-J,J]$. The transition with $\Delta N_J=\sigma$, which corresponds to the spin-only transfer to the molecular (internal) rotation, is few orders of magnitude stronger than the transitions, related to the OAM transfer. Mixing these transitions in spectrum renders it impossible to distinguish between $M=0$ and $M\neq 0$ beams. The straightforward solution is to lift the degeneracy of molecular energies with respect to $N_J$. For instance, this can be done by applying a strong static electric field $E_\text{Stark}$ along the beam axis. Such a field induces the AC Stark shift of ro-vibrational energy levels with different values of $| \Delta N_J|$. The ro-vibrational energies of the molecule in the presence of $E_\text{Stark}$ can be calculated by diagonalizing the Hamiltonian: $\mathcal{H}_\text{mol}=\mathcal{H}_\text{Dun}+\mathcal{H}_\text{Stark}$, where $\mathcal{H}_\text{Dun}$ is the Dunham expansion Hamiltonian, which is diagonal with respect to molecular rotational states $\ket{J,N_J}$, and $\mathcal{H}_\text{Stark}=d_z E_\text{Stark}C^{J^\prime,N_J}_{J,N_J;1,0}\vert J^\prime N_J\rangle\langle J N_J\vert$, where $C^{J^\prime,N_J}_{J,N_J;1,0}$ are the Clebsch-Gordan coefficients~\cite{Varshalovich} and $d_z$ is the equilibrium dipole moment at the given vibrational level.

We present the absorbance spectrum in Fig.~\ref{fig_experiment}(c). The positions of spectral lines that correspond to ro-vibrational transitions with $\Delta \nu=1$ and $\Delta J=2$, associated with the electric-quadrupole interaction~\eqref{eq_Hamiltonian_quad}, are marked by the dotted lines. Out of three peaks for $|\Delta N_J|=\{0,1,2\}$, we observe two: the higher peak corresponds to the spin-only transition with $\Delta N_J=\sigma$ and the lower peak corresponds to the OAM transfer to the molecular (internal) rotation. As expected, the OAM-enabled peak is strongly suppressed for the Gaussian beam with zero vorticity $(M=0)$. Another peak, associated with the OAM transfer to molecular (internal) rotation (with $|\Delta N_J|=0$), is not enhanced, as discussed in Sec.~\ref{sec_LG_beams}, and is therefore indistinguishable in the vicinity of the strong $|\Delta N_J|=1$ transition. The nearest $\Delta J=1$ transition, associated with the electric-dipole interaction~\eqref{eq_Hamiltonian_dip}, is marked with arrow. Even though, it is fully saturated (with $\mathcal{A}\approx 1$), it is sufficiently distant not to affect the signal from quadrupole transitions. This large splitting between the dipole and quadrupole transitions is a manifestation of the ro-vibrational coupling, described by the Dunham expansion. To achieve a visible splitting between spectral lines with different $|\Delta N_J|$ values, we choose the Stark field amplitude $E_\text{Stark}=7\,\text{kV}/\text{cm}$. In the inset of Fig.~\ref{fig_experiment}(c), we plot the ratio between the splitting of $|\Delta N_J|=1$ and $|\Delta N_J|=2$ spectral lines, induced by the AC Stark shift, and the Doppler-broadening of the $|\Delta N_J|=1$ transition, as the function of the electric field magnitude $E_\text{Stark}$. Different curves correspond to commonly used cryogenic temperatures: of liquid \mbox{He~(4.222 K),} Ne~(27.1 K) and N$_2$~(77.36 K)~\cite{CryoTemp}.

Figure~\ref{fig_experiment}(c) reveals that the vorticity-induced changes to the ro-vibrational spectrum are already traceable in a rather simple proof-of-principle measurement. Of course, some aspects of the suggested experimental scheme can be optimized. In particular, our numerical analysis reveals several ways to improve a possible experimental protocol:
\begin{enumerate}
\item \textit{Choice of the molecule.} The absorbance values of $\Delta N_J\neq \sigma$ spectral lines are proportional to squares of vibrational transition quadrupole moments, as revealed by Eqs.~\eqref{eq_transition_matrix} and~\eqref{eq_Fermi_rule}. At the same time, the AC Stark splitting between different $|\Delta N_J|$ lines is quadratic in terms of the equilibrium dipole moment. This implies, that one should search for the molecule or, specifically, for the particular ro-vibrational transition, that enables the right balance between high equilibrium dipole and high transition quadrupole moments.

\item\textit{Length ratios.} As discussed in Sec.~\ref{sec_LG_beams}, the rotational enhancement occurs close to the axial focal point of the beam $(|R|\leq \omega_0,|Z|\leq Z_R)$. The interaction of light with molecules outside this domain does not display the enhanced absorbance. To achieve the optimal spectral signal, one should target $L\approx Z_R$ and $R_0<\omega_0$. We do not have the definitive suggestion on the choice of geometry, since the lengths are interconnected with one another as well as other parameters of the system. For instance, a lower beam waist $\omega_0$ would imply a lower Rayleigh range $Z_R$ and, thus, necessitate using a shorter chamber. Also, although a smaller aperture (compared to the beam waist) result in higher absorbance values, it also implies that less photons are participating in the interaction, since the vortex beam has a dark core. Nevertheless, our approach can be used to optimize the geometric parameters in a case by case basis by considering the particular limitations of the experimental apparatus.

\item\textit{Pressure.} The effect of the pressure in the gas chamber on the absorbance values is ambivalent. On the one hand, increasing the pressure would increase the number of absorbers (molecules) within the optical path. On the other hand, higher pressure would result in the sizeable pressure broadening of the spectral lines, which, at some point, may become the dominant broadening mechanism.

\item\textit{Temperature.} Unlike the other parameters, the effect of temperature on the spectrum is unambiguous. It is beneficial to have a lower temperature. This would result in lower center-of-mass velocities as well as the weaker Doppler broadening. Ideally, one could even consider optical trapping and laser cooling of molecules~\cite{Anderegg_2018,Mitra_2020,Zhang_2020,Yu_2021,Cairncross_2021,Langen_2023,Koch_2019}.
\end{enumerate}

Apart from the ro-vibrational spectroscopy of molecules, there may be other effective methods to measure the enhancement of the OAM transfer to the internal rotation of a particle, induced by the non-zero helicity of the beam. Some of these methods, like levitated optomechanics, are discussed in App. \ref{app_extensions}.

\section{Conclusion and outlook}
\label{sec_conclusion}

Motivated by the ubiquity of light-matter interactions in modern-day physics, as well as the multifaceted nature of the angular momentum of light, we developed a general analytical framework that describes the angular momentum exchange associated with the interaction of particles with the spin and OAM of light. To keep our explanation concise and illustrative, while introducing the general idea, we focused on the ro-vibrational spectroscopy of molecules. This is a prototypical system, where the emergent angular momentum exchange is encapsulated by the selection rules, which have direct experimental signatures.

Our analytical framework is based on the multipole expansion of the light-matter interaction. On the example of the rotation and vibrations of a molecule, we explain how internal degrees of freedom of a particle can be embedded into the definition of its multipole moments. Since our main focus is on the AM exchange that is associated with the light-matter interaction, we restrict our model to molecules without strong ro-vibrational coupling. However, the general idea behind our model can be applied to a broad range of systems, where the AM exchange is of the main interest, by introducing the additional couplings of rotations to the relevant degrees of freedom (see App.~\ref{app_extensions}).

Our derivation of the light-matter interaction Hamiltonian~\eqref{eq_interaction_Hamiltonian_final} takes full advantage of the notions of rotations and angular momentum. In particular, we describe the molecule using spherical multipole moments, which reflect its (symmetry) point group, and expand the electric field in terms of its spherical gradients. As a result, our theory unambiguously describes the angular momentum exchange in the case of the ro-vibrational spectroscopy. Our treatment is conceptually similar to the Wigner-Eckart theorem. It allows us to circumvent the detailed analysis of the molecular and optical field structure, by absorbing their specifics into the definition of the spherical expansion coefficients. \textit{This is the definitive feature of our framework.} Previous studies on ro-vibrational spectroscopy using vortex beams~\cite{Babiker_2002,Alexandrescu_2006,Mondal_2014,Mukherjee_2018} addressed the full problem of the light-matter interaction, which required introducing restrictive models of either molecule or optical field.

To demonstrate the predictive power of our theory, we calculated the selection rules for the ro-vibrational spectroscopy using Laguerre-Gaussian beams. We confirmed a well-known result~\cite{Babiker_2002,Mondal_2014,Andrews_2004} that, for non-tightly focused beams within the electric-dipole interaction, the OAM of light cannot couple to the molecular (internal) rotation irrespective of the beam profile. At the quadrupole order of the interaction, however, we demonstrate that the helicity of the beam strongly enhances certain ro-vibrational transitions, which are considered ``forbidden'' in the case of  non-twisted light. The enhancement strongly depends on the mean position of the molecular center of mass with respect to the focal point of the beam. Recent years have seen a rapid development in trapping, cooling and quantum control of molecules~\cite{Anderegg_2018,Mitra_2020,Zhang_2020,Yu_2021,Cairncross_2021,Langen_2023,Koch_2019}. Therefore, it is realistic to suggest measuring the ro-vibrational enhancement in these new setups. Another pathway is to study the gas-phase absorption spectroscopy in a spatially and frequency-resolved manner. We provide proof-of-principle calculations by suggesting an experiment that may be capable of probing the enhancement. Based on our findings, we discuss the experimental requirements for the observation of the enhanced ro-vibrational transitions.

Being straightforward and general, our theory opens new research avenues for the studies that involve angular momentum exchange induced by light-matter interaction. Its applications can reach beyond ro-vibrational spectroscopy. 
In particular, our analysis establishes four main actors of the interaction process. Namely, the spin and the OAM of light, as well as, the AM-related (rotational) and the AM-unrelated (e.g., vibrational) degrees of freedom of a particle. The scenario, where the intra-light and intra-matter couplings are negligible, can be described by our perturbative argumentation. Nevertheless, the introduction of either of these couplings can lead to richer physics, as it introduces a competition between the different interaction pathways.

Apart from the pure theoretical interest, studies of AM exchange in the presence of additional couplings are highly relevant for the experiment. Considering intra-light interactions, tight focusing was shown to benefit the OAM interaction with chiral molecules and nanostructures~\cite{Wozniak_2019,Forbes_2021}. In such electric fields, the spin and OAM of light are strongly coupled~\cite{Zhao_2007,Bliokh_SOC}, which enables OAM transfer already at the dipole order of the light-matter interaction. Hence, a detailed analysis of such systems within our framework is a promising endeavor. On a related note, the first step towards understanding the AM transfer in systems with more complicated material response, concerns the study of the helical dichroism~\cite{Rouxel_2022,Kerber_2018,Forbes_2018,Forbes_2019,Forbes_2021,Brullot_2016,Begin_2023} -- a novel experimental technique for enantiomer resolution. The current understanding of this phenomenon revolves around the interplay of different multipolar orders of the light-matter interaction~\cite{Andrews_review_2018}, which can be enabled by the ro-vibrational interaction. Hence, analyzing the AM exchange in the presence of strong ro-vibrational coupling can outline the efficacy of the helical dichroism for enantiomer resolution. Finally, in addition to the electric multipole interactions, chirality-related physics was also argued to involve the magnetic multipoles. Their interaction with magnetic fields can also be expressed in terms of a model Hamiltonian, similar to our theory, by defining an effective magnetization density for the molecule.

\section*{Acknowledgements}

We are grateful to Emilio Pisanty and Philipp Lunt for valuable discussions. This research was funded in whole or in part by the Austrian Science Fund (FWF) [10.55776/F1004]. G.M.K. gratefully acknowledges funding from the European Union’s Horizon 2020 research and innovation programme under the Marie Skłodowska-Curie Grant Agreement No.~101034413. M.L. acknowledges support by the European Research Council (ERC) Starting Grant No.~801770 (ANGULON). O.H.H. acknowledges support by the Austrian Science Fund (FWF) [10.55776/P36040]. Furthermore, the financial support by the Austrian Federal Ministry for Digital and Economic Affairs, the National Foundation for Research, Technology and Development and the Christian Doppler Research Association is gratefully acknowledged.

\appendix

\section{Extensions to the multipole model}
\label{app_extensions}

The effective model of molecular multipole moments, introduced in Sec.~\ref{sec_multipole_moments}, describes only the dependence of multipoles on the molecular rotation and vibrations. Our particular suggestion for the charge distribution~\eqref{eq_charge_density_mol} is not capable of describing other degrees of freedom or other particle types. However, the underlying principle can be applied to other systems, where the angular momentum exchange between the particle and surrounding electric field is of the main interest. Here, we discuss possible extensions to our model.

\subsection{Ro-vibrational coupling}

Instead of resorting to numerical methods to calculate vibrational transition multipole moments, as in Sec.~\ref{sec_multipole_moments}, one could introduce an analytical model of vibrations to reveal more information about the coupling mechanism between different excitations. 

The na\"ive approach is to model the vibrational dependence of multipoles with the modes of the quantum harmonic oscillator. If one considers a molecule with a single vibrational degree of freedom, like  a diatomic molecule, multipole moments can be expanded in Taylor series with respect to the vibrational coordinate $\hat{q}$. For instance, for the quadrupole moment matrix $Q_{i,j}$, the expansion reads
\begin{align}
    Q_{i,j}(\hat{q})&=Q_{i,j}(\hat{q}=0)+\bigg[\frac{\partial Q_{i,j}}{\partial q}\bigg]_{\hat{q}=0}\hat{q}+...\nonumber\\
    &=\sum\limits_{\nu,\nu' = 0}^{\infty} Q_{i,j}^{\nu,\nu'}(\hat{a}^\dagger)^{\nu'}(\hat{a})^{\nu}\,,
    \label{eq_vibration}
\end{align}
where $i,j\in\{x,y,z\}$ and $\hat{q}=q_0(\hat{a}^\dagger+\hat{a})$, where $\hat{a}^{(\dagger)}$ is the (creation) annihilation operator of the vibrational mode and $q_0$ is the characteristic distance. \mbox{The coefficient $Q_{i,j}^{\nu,\nu'}$} is the quadrupole moment matrix associated with the transition from the vibrational state~$\ket{\nu}$ to the state $\ket{\nu'}$. In Sec.~\ref{sec_multipole_moments}, we show that, in order to simulate the specific molecule, matrices~$Q_{i,j}^{\nu,\nu'}$ could be extracted from numerical ab-initio calculations. Alternatively, these matrix elements can be considered as free parameters of the model and used to study its emergent physics. In this case, it is important to explicitly constrain the multipole moments in a way that reflects the molecular (symmetry) point group~$\mathcal{G}_\text{mol}$~(see App.~\ref{app_molecular_rotation_group}). This approach can be straightforwardly generalized to the case of multiple contributing vibrational modes $\hat{q}_n$.

The main downside of the na\"ive model is that it does not allow for the coupling between the molecular rotation and its vibrational modes. In other words, it requires the ground and all excited vibrational states to be characterized by the same molecular point group~$\mathcal{G}_\text{mol}$, which is not true for an arbitrary molecule. In situations, where the ro-vibrational coupling is sizeable, it is necessary to consider a more accurate model than the na\"ive oscillator model. For instance, one can work with the molecule in the Eckart reference frame~\cite{Louck_1976}, which helps to approximately separate vibrations and rotations. Although, the full separation is impossible, Eckart conditions help to minimize the ro-vibrational coupling in the equillibrium configuration. In this reference frame, Watson Hamiltonian provides the complete description of the rotation and vibrations of a linear~\cite{Watson_1993} or non-linear~\cite{Watson_1968} molecule, omitting excitations of their electronic structure.

\subsection{Molecular electronic transitions}

The multipole model from Sec.~\ref{sec_multipole_moments} is adiabatic with respect to molecular electronic transitions. This is justified by our focus on the ro-vibrational spectroscopy of molecules. In such experiments, the time, required for the reshaping of the electronic cloud, is much shorter than the characteristic time of the rotation or vibration of the molecular nuclear backbone. Thus, one can assume that electrons readjust instantaneously during molecular transitions. In contrast, when one directly probes or drives electronic transitions, e.g., in vibronic spectroscopy~\cite{Herzberg_book}, one needs to account for the coupling of electronic transitions to other degrees of freedom. In particular, one can no longer consider only vibrational transitions within a single potential energy surface, like we did in Sec.~\ref{sec_multipole_moments}. Instead, one needs to calculate the multipole moments associated with transitions between vibrational levels of two (or more) different potential energy surfaces. Apart from this vibronic coupling, electronic transitions are also capable of changing the molecular \mbox{(symmetry)} point group $\mathcal{G}_\text{mol}$. This should be reflected in the definition of the molecular orientation~$\hat{\Omega}$ and spherical multipole moments $\alpha_{\lambda,\mu}$ \mbox{(see App. ~\ref{app_molecular_rotation_group}).} For instance, in the case of a homonuclear diatomic molecule, which has zero dipole moment at equilibrium, i.e., electronic ground state, it is possible to induce the non-zero dipole moment by exciting the electronic cloud around the molecule.

\subsection{Electronic orbitals in atoms}

In contrast to the molecular case, it is not possible to define the orientation of an atom at equilibrium. As a result, it is not possible to introduce the reference frame, co-rotating with the atom, which renders the charge distribution~\eqref{eq_charge_density_mol} unusable. Nonetheless, it may still be possible to apply our theory to the electronic orbitals. Energy levels of the electronic cloud around the atom can be described by spectroscopic term symbols~\cite{Corney_book}, which are sets of ``good'' quantum numbers, namely the spin, orbital and total angular momenta of atom's electrons. The Wigner-Eckart theorem~\cite{Hall_Lie} states that a matrix element, associated with a transition between two energy levels of an atom, can be always expressed as a product of two factors. The first factor is the reduced matrix element, which is independent of the orientation of angular momenta and is the same for all transitions between energy levels, characterized by the same term symbol. The second factor is the Clebsch-Gordan coefficient that encapsulates the angular momentum exchange during the transition. Owing to the Wigner-Eckart theorem, our model can be extended to the atomic case, by defining an ``effective atomic frame'', which depends on the initial and final term symbols and reproduces the decomposition of the transition matrix elements. Inside this reference frame, one can introduce a charge distribution, similar to Eq.~\eqref{eq_charge_density_mol}, and perform the analysis in a way, similar to Sec.~\ref{sec_multipole_moments}. 

%Atoms at equilibrium are in principle spherically symmetric objects owing to the symmetry of the Coulomb potential, and thus the concept of atomic rotation does not make sense. However, when atoms are subjected to external fields this symmetry can be reduced. As a concrete example consider the case of a $s \to p$ dipole transition, which can be characterized as $\pi_{\pm}$ or $\sigma$ depending on the projection of the angular momentum of the final state on the ``quantization axis'' given by the lab-frame field configuration. Our findings can be generalized to also account for atoms by identifying the spherical multipole moments $a_{\lambda, \mu}$ in terms of the reduced matrix elements via the Wigner-Eckart theorem and considering the rotations of the effective charge distribution as giving rise to the corresponding Clebsch-Gordan coefficients.

\subsection{Chiral particles}

From the geometric standpoint, optical beams with the spiral beam phase, like Laguerre-Gaussian modes, discussed in Sec.~\ref{sec_LG_beams}, are chiral, as they lack the axis of an improper rotation. Similar to other chiral electric fields~\cite{Ayuso_2019,Vogwell_2023,Mayer_2023}, vortex beams were argued to be a natural tool for the molecular enantiomer resolution~\cite{Rouxel_2022}. In particular, the interplay between the optical and material chirality is subject to the dichroism~\cite{Andrews_review_2018}, which is manifested by the dissimilarity of an optical activity indicator, like absorbance, for different values of the ``handedness'' of the field or particle.

The dichroism associated with the change to the OAM of light is often called \textit{helical dichroism}~\cite{Rouxel_2022,Kerber_2018}. This effect was proven to be absent in the dipole order of the light-matter interaction~\cite{Andrews_2004,Araoka_2005}, but it was argued to be observable either using electric quadrupole fields~\cite{Brullot_2016,Forbes_2018,Forbes_2019} or tightly-focused laser beams~\cite{Wozniak_2019,Forbes_2021}. Moreover, the magnitude of the effect was shown to depend on the position of the phase singularity~\cite{Begin_2023}, similar to the OAM transfer that is discussed in Sec.~\ref{sec_LG_beams}. Formally, one of the possible mechanisms for the helical dichroism is the so-called E1-E2 interaction. It implies driving the (molecular) transition, which is associated with the simultaneously non-zero values of (transition) dipole and quadrupole moments. Such transitions are known to be found between the molecular states with an ill-defined AM parity~\cite{Kral_2007}. Mixed-parity states cannot be derived using a rigid rotor model (see App.~\ref{app_molecular_rotation_group}), since the Hamiltonian of a rigid rotor commutes with the molecular AM operator $\hat{J}^2$, which necessitates every rotational state to have a well-defined parity ($J$).

It may be possible to describe the mixed-parity molecular states, if one takes proper care of the ro-vibrational coupling, e.g., by using the Watson Hamiltonian~\cite{Watson_1968,Watson_1993}. Alternatively, one can consider other effective models, for instance, based on \textit{local currents}~\cite{Fernandez_Corbaton_2017}, which might give rise to a more intuitive picture of the light-matter interaction. Moreover, (effective) currents may be instrumental for the calculation of torroidal multipole moments, which were shown to be associated with the order parameters that describe chirality~\cite{Kishine_2022}.

\subsection{Optically levitated nanoparticles}

The last years have seen a tremendous improvement in the field of levitated optomechanics~\cite{Stickler_2021}. Optically levitated nanoparticles, like silicon nanorods, are among the highest quality sensors for various forces and fields. They can be controlled with an exquisite precision in an experiment~\cite{Kuhn_2017}: their dimensions can be tailored, allowing for the high degree of reproducibility, and their orientation and rotation can be controlled using using the radiation pressure. Such nanorotors were also shown to interact with the OAM of light~\cite{Hu_2023}, which makes them a possible experimental platform for benchmarking our theory.  In particular, it might be possible to use these sensitive nanorotors to measure the dependence of the OAM transfer on the position of the center of mass, thus experimentally verifying the findings of Sec.~\ref{sec_LG_beams}.

In addition to the experiment, the theoretical analysis of nanoscale rotating objects is also of a fundamental interest. For instance, Ref.~\cite{Ma_2020} suggests that quantum rotations of thermal asymmetric nanorotors exhibit a more robust coherent flipping dynamics, known as the tennis racket effect, than their classical analogues. Contrasted with molecules, the characteristic sizes of nanoparticles can be comparable to the wavelength of light. Although, in this scenario the multipole expansion still holds, if the multipole moments are properly defined~\cite{Alaee_2018}, the magnetic multipole moments of a particle may become relevant for describing the light-matter interaction. In this case, alongside with the charge distribution~\eqref{eq_charge_density_mol}, discussed in Sec.~\ref{sec_multipole_moments}, one would need to define a \textit{distribution of local currents} that reflects the dependence of magnetic multipoles on internal degrees of freedom.

\section{Molecular rotation group}
\label{app_molecular_rotation_group}

Defining the orientation of a \textit{completely asymmetric} rigid rotor in three dimensions is straightforward. Rotations of such an object form the SO(3) group~\cite{Wigner_book}, known simply as the rotation group. Each element of this continuous group, denoted as the rotation operator~$\hat{\mathcal{D}}(\Omega_\text{rot})$, can be parameterized by three Euler angles $\Omega_\text{rot}\equiv\{\alpha,\beta,\gamma\}$. Euler angles $\Omega_\text{rot}$ uniquely define the orientation of a rotor, except for the case of $\beta=0$, known as the problem of a gimbal lock.

The action of the rotation operator $\hat{\mathcal{D}}(\Omega_\text{rot})$ on the Euclidean space~$\mathbb{R}^3$ is determined by the irreducible representations (or irreps) of the SO(3) group. The representation of an odd degree~$l$ is an orthogonal set of $l$ matrices, called Wigner D-matrices $D^{l}_{m,n}(\Omega_\text{rot})$~\cite{Varshalovich}, defined in the $(2l+1)$-dimensional vector space~$\mathfrak{H}_l$. This vector space is the SO(3)-invariant subspace of the linear space $\mathcal{P}_l$ of all homogeneous polynomials of \mbox{the degree $l$ on $\mathbb{R}^3$~\cite{Broecker_2013}.} Therefore, it is irreducible. The space~$\mathfrak{H}_l$ is spanned by the spherical harmonics $Y_{l,m}(\Omega)$, with $\Omega\in \text{S}_2$. The action of the group element~$\hat{\mathcal{D}}(\Omega_\text{rot})$ on the spherical harmonic~$Y_{l,m}(\Omega)$ is defined by Eq.~\eqref{eq_sph_harm_rotation}.

Besides serving as irreducible representations of the rotation group, Wigner matrices $D^{l}_{m,n}(\Omega_\text{rot})$ also function as the Fourier transform, due to the Pontryagin duality~\cite{Broecker_2013}. For a rigid rotor, it is common to consider two bases: the basis of Euler angles $\ket{\Omega_\text{rot}}$, defined on the SO(3) manifold as the eigenbasis of the orientation operator $\hat{\Omega}_\text{rot}$, and the conjugate basis of angular momenta $\ket{l,m,n}$, defined on the tangent space $\widehat{\text{SO(3)}}$ as the simultaneous eigenbasis of the laboratory-frame $\hat{\bm{L}}^2$ and $\hat{L}_z$ and body-fixed-frame $\hat{L}'_z$ angular momentum operators, respectively. The Fourier transform between the two bases reads
\begin{equation}
    \ket{\Omega_\text{rot}}=\sum\limits_{\{l,m,n\}\in \widehat{\text{SO(3)}}}\sqrt{\frac{2l+1}{8\pi^2}}D^{l,\ast}_{m,n}(\Omega_\text{rot})\ket{l,m,n}\,.
    \label{eq_fourier_transform_rot_bas}
\end{equation}
The inverse of this transform is commonly used in quantum mechanics to define the wave functions with a given angular momentum in the basis of Euler angles.

Defining the orientation of a \textit{molecule} $\Omega_\text{mol}$ is different from the case of the rigid rotor. Molecules are usually characterized by a certain (symmetry) point group~$\mathcal{G}_\text{mol}$. Therefore, their rotations comprise only a subgroup of the full rotation group. Namely, they form the quotient group: $\mathcal{Q}\equiv \text{SO(3)}/\mathcal{G}_\text{mol}$~\cite{Albert_2020}. For instance, for the linear molecule, like CS: $\mathcal{G}_\text{mol}=\text{C}_{\infty}=\text{U}_1$ and the quotient group is a two-sphere $S_2=\,$SO(3)$/\text{U}_1$. This can be understood intuitively. Unlike the rigid rotor case, where all three Euler angles $\Omega_\text{rot}\equiv\{\alpha,\beta,\gamma\}$ are required to define the orientation, for a linear molecule, which is symmetric with respect to the rotation around the molecular axis, only two angles $\Omega_\text{mol}\equiv\{\beta,\alpha\}\in \text{S}_2$ are sufficient.

Irreducible representations of the quotient group $\mathcal{Q}$ are integrals of the irreps of the SO(3) group with respect to the point group $\mathcal{G}_\text{mol}$. In the case of a linear molecule, the representations are spherical harmonics
\begin{equation}
    D^{l}_{m,n}(\Omega_\text{mol})=\int\limits_{U_1}D^{l}_{m,n}(\alpha,\beta,\gamma)\frac{\mathrm{d}\gamma}{|\text{U}_1|}\propto\delta_{n,0}Y^{\ast}_{l,m}(\Omega_\text{mol})\,,
    \label{eq_irrep_molecule}
\end{equation}
where $|\text{U}_1|=2\pi$ is the volume of the group U$_1$ as a manifold. Similar to Eq.~\eqref{eq_fourier_transform_rot_bas}, irreducible representations~\eqref{eq_irrep_molecule} serve as the Fourier transform between the molecular angle basis $\ket{\Omega_\text{mol}}$ and the conjugate angular momentum basis, defined on $\widehat{\mathcal{Q}}$. For a linear molecule, the angular momentum eigenstates are denoted as $\ket{l,m}$, since $\hat{L}'_z\equiv 0$ renders index $n$ irrelevant.

If the rotational operator $\hat{\mathcal{D}}(\Omega_\text{mol})$ is being applied to the charge distribution~\eqref{eq_charge_density_mol}, instead of the correct representations of the quotient group $\mathcal{Q}$, one can simply use the representations of the full SO(3) group. The values of the spherical multipole moments $\alpha_{\lambda,\mu}$ already embed the molecular (symmetry) point group $\mathcal{G}_\text{mol}$~\cite{Gelessus_1995}. Therefore, if the transformation~\eqref{eq_sph_harm_rotation} with incorrect irreducible representations was to generate any excess terms, they would be removed by the vanishing multipole moments associated with them.

Internal molecular degrees of freedom are capable of changing the molecular (symmetry) point group $\mathcal{G}_\text{mol}$ by effectively coupling to rotations. For instance, in the case of a polyatomic linear molecule, like OCS, vibrational bending modes or molecular electronic transitions can break the U$_1$ symmetry, which characterizes the molecule at equilibrium. This fact necessitates introducing different definitions $\Omega^{g(e)}_\text{mol}$ for the orientation of the molecule in the ground and excited vibrational or electronic states. Subsequently, the rotation rules~\eqref{eq_sph_harm_rotation}, which define the transformation into the body-fixed frame, are no longer the same for the molecule at and outside equilibrium. As a result, for polyatomic molecules and other types of particles, it is not possible to simply detach rotations from other excitations, like we did in Sec.~\ref{sec_multipole_moments} for molecular vibrations. In such cases, a more accurate model should be introduced. For instance, in the case of vibrations, one could consider a Hamiltonian that includes the ro-vibrational coupling (see App.~\ref{app_extensions}), like Watson Hamiltonian~\cite{Watson_1968,Watson_1993},  together with the light-matter Hamiltonian~\eqref{eq_interaction_Hamiltonian_final}. In this case, the molecular frame should be defined so that it properly accounts for the configuration of lowest symmetry.

\section{Derivation of the light-matter interaction Hamiltonian}
\label{app_hamiltonian}

We consider the light-matter interaction part of the Power-Zienau-Wooley Hamiltonian~\cite{Power_1959,Atkins_1970,Wooley_1971} after the generalization to the case of a continuous charge distribution~$\rho(\bm{r})$,~i.e.,~we begin with Eq.~\eqref{interaction_Hamiltonian_integral}. We expand the spatial electric field profile~$\bm{E}(\bm{R}+\eta\bm{r})$ into the spherical Taylor series~\cite{Weniger_2002,Weniger_2000,Weniger_2005} around the center-of-mass position~$\bm{R}$, using Eq.~\eqref{eq_Taylor}. As a result, we obtain the following expression for the interaction Hamiltonian:
\begin{align}
    &\mathcal{H}_{\text{int}}=-\sqrt{\frac{4\pi}{3}}\sum\limits_{n,l,m,\sigma}\frac{c_{n,l}}{2n+l+1}\big[\mathcal{R}_{l,m}(\bm{\nabla}_{\bm{R}}) E_\sigma(\bm{R})\big]\nonumber\\
    &\times\lim_{\chi\to 0}\int\mathrm{d}^3\bm{r}\,r^{2n+l+1}\rho(\bm{r})Y_{1,\sigma}(\Omega_{\bm{r}})Y^{\ast}_{l,m}(\Omega_{\bm{r}})+\text{H.c.}\,,
    \label{eq_app_c_1}
\end{align}
where the scalar product is evaluated using the spherical expansion: $(\bm{A}\cdot \bm{B})=\sqrt{4\pi/3}\,|A|\sum_{\sigma}B_\sigma Y_{1\sigma}(\Omega_{\bm{A}})$, where $\bm{A}=\{|A|,\Omega_{\bm{A}}\}$ and $\bm{B}$ are vectors in the laboratory frame, and ${B_\sigma\equiv\{B_{\pm},B_0\}=\{(B_{x}\pm iB_{y})/\sqrt{2},B_{z}\}}$ are the components of vector $\bm{B}$ in the spherical basis. Spherical components $E_\sigma(\bm{R})$ of the electric field vector $\bm{E}(\bm{R})$ are often called \textit{circular polarization components}.

We further substitute $\rho(\bm{r})$ in Eq.~\eqref{eq_app_c_1} with the generic charge distribution~\eqref{eq_charge_density_mol} of a system of multipoles. Note that the charge distribution~\eqref{eq_charge_density_mol} is defined in the rotating reference frame, which is uniquely characterized by the orientation $\hat{\Omega}$, discussed in App.~\ref{app_molecular_rotation_group}. The rotating-frame coordinate $\bm{r}'$ is a function of both the laboratory-frame coordinate $\bm{r}$ and the orientation $\hat{\Omega}$, i.e., $\bm{r}'\equiv\bm{r}'(\bm{r},\hat{\Omega})$. Since $|\bm{r}|=|\bm{r}'|$, one can easily evaluate the radial integral over $r\equiv|\bm{r}|$ and the limit $\chi\to0$, obtaining
\begin{align}
    \lim_{\chi\to 0}&\int\mathrm{d}^3\bm{r}\,r^{2n+l+1}\rho(\bm{r})Y_{1,\sigma}(\Omega_{\bm{r}})Y^{\ast}_{l,m}(\Omega_{\bm{r}})=\nonumber\\
    &\sum\limits_{\lambda,\mu}\hat{\alpha}_{\lambda,\mu}\int\mathrm{d}\Omega_{\bm{r}}\,\mathcal{Y}_{\lambda,\mu}(\Omega_{\bm{r}'})Y_{1,\sigma}(\Omega_{\bm{r}})Y^{\ast}_{l,m}(\Omega_{\bm{r}})\,,
    \label{eq_app_integral}
\end{align}
with the additional condition $2n+l-\lambda+1\leq 0$ such that the limit is non-vanishing. To evaluate the integral over the solid angle~$\Omega_{\bm{r}}$, one needs to express the molecular-frame real-valued harmonics~$\mathcal{Y}_{\lambda,\mu}(\Omega_{\bm{r}'})$ through the laboratory-frame complex-valued harmonics~$Y_{\lambda,\zeta}(\Omega_{\bm{r}})$, with the help of \mbox{Eqs.~\eqref{eq_real_harmonics} and \eqref{eq_sph_harm_rotation}.} The result reads:
\begin{align}
    &\mathcal{Y}_{\lambda,\mu}(\Omega_{\bm{r}'})=\hat{\mathcal{D}}(\hat{\Omega})\mathcal{Y}_{\lambda,\mu}(\Omega_{\bm{r}})=\frac{1}{\sqrt{2}}\sum\limits_{\zeta}\nonumber\\
    &\times\begin{cases}
    i\big(D^{\lambda}_{\zeta,\mu}(\hat{\Omega})Y_{\lambda,\zeta}(\Omega_{\bm{r}})-D^{\lambda\ast}_{\zeta,\mu}(\hat{\Omega})Y^\ast_{\lambda,\zeta}(\Omega_{\bm{r}})\big)&\mu<0\\
    \sqrt{2}D^{\lambda}_{\zeta,0}(\hat{\Omega})Y_{\lambda,\zeta}(\Omega_{\bm{r}})&\mu=0\\
    D^{\lambda}_{\zeta,-\mu}(\hat{\Omega})Y_{\lambda,\zeta}(\Omega_{\bm{r}})+D^{\lambda\ast}_{\zeta,-\mu}(\hat{\Omega})Y^\ast_{\lambda,\zeta}(\Omega_{\bm{r}})&\mu>0\end{cases}\,,
\end{align}
Finally, the angular integral in Eq.~\eqref{eq_app_integral} is evaluated using~\cite{Varshalovich}
\begin{align}
    \int\mathrm{d}\Omega_{\bm{r}}\,Y_{\lambda,\zeta}(\Omega_{\bm{r}})&Y_{1,\sigma}(\Omega_{\bm{r}})Y^{\ast}_{l,m}(\Omega_{\bm{r}})\nonumber\\
    &=\sqrt{\frac{3(2\lambda+1)}{4\pi(2l+1)}}C^{l,0}_{\lambda,0;1,0}C^{l,m}_{\lambda,\zeta;1,\sigma}\,,
\end{align}
where $\lambda\leq l+1$ is required for the Clebsch-Gordan coefficients to be non-zero. This condition, together with the condition from Eq.~\eqref{eq_app_integral}, implies that $n=0$.

\section{Approximations to the Laguerre-Gaussian profile}
\label{app_EF_approximations}

The electric field profile $E_{P,M}(\bm{R})$ of a Laguerre-Gaussian mode, defined in Eq.~\eqref{eq_EF_Allen}, is a convoluted function of cylindrical coordinates $\bm{R}\equiv\{R,\Phi,Z\}$. While it is possible to calculate spherical gradients~\eqref{eq_spherical_gradient} of $E_{P,M}(\bm{R})$ analytically, the resulting expressions are not straightforward to analyze. Thus, in Eq.~\eqref{eq_EF_approximation_1}, we define the approximation~$E^{\text{foc}}_M(\bm{R})$ to the full profile~$E_{P,M}(\bm{R})$. This expression corresponds to a LG beam with ${P=0}$ within the \textit{in-focus approximation}. It accurately describes the electric field~$E_{P,M}(\bm{R})$ inside the Rayleigh range $(Z\ll Z_{R})$. Previous studies on the interaction of molecules with vortex beams~\cite{Alexandrescu_2006,Mondal_2014} have already introduced a similar approximation
\begin{equation}
    %E^{\text{foc,cent}}_M(\bm{R})=1/\sqrt{|M|!} (R/\omega_0)^{|M|}e^{-iM\Phi}e^{-ikZ}\,,
    E^{\text{foc,cent}}_M(\bm{R})=\frac{1}{\sqrt{|M|!}} \left(\frac{R}{\omega_0}\right)^{|M|}e^{-iM\Phi}e^{-ikZ}\,,
    \label{eq_EF_approximation_2}
\end{equation}
which corresponds to Eq.~\eqref{eq_EF_approximation_1} with the additional \textit{in-center approximation}. It is valid inside the beam waist close to the beam axis $(R\ll \omega_0)$.

Even though Eq.~\eqref{eq_EF_approximation_2} includes both major features of the LG profile, namely, the spiral beam phase and the radial magnitude scaling near the beam center, the electric field~$E^{\text{foc,cent}}_M(\bm{R})$ diverges at large distances $R$. While this problem can be solved by introducing a radial cutoff $\omega_{\text{cut}}\sim\omega_0$, sometimes called the impact parameter, the value of this cutoff was argued to affect the results of numerical calculations~\cite{Wu_2022}. Moreover, the light-matter interaction was shown to depend on the average position of molecule's center of mass with respect to the cutoff distance~$\omega_{\text{cut}}$~\cite{Mondal_2014}. Therefore, to ensure the accuracy of our findings, we choose the cutoff-free approximation~\eqref{eq_EF_approximation_1} to the electric field~\eqref{eq_EF_Allen}.

Another reason to use the approximation~\eqref{eq_EF_approximation_1} over the approximation~\eqref{eq_EF_approximation_2} becomes apparent after the calculation of the spherical gradient~\eqref{eq_spherical_gradient}. In the case of $E^{\text{foc,cent}}_M(\bm{R})$, the term in Eq.~\eqref{eq_gradient} associated with the radial exponential decay of $E_{P,M}(\bm{R})$ is absent from the expression for the gradient. Hence, for the Laguerre-Gaussian beam with $M=0$, the spherical gradient and, thus, the magnitude of the OAM transfer is zero, as also shown in Ref.~\cite{Mondal_2014}. The possible conclusion is that a non-helical beam cannot transfer any additional angular momentum, apart from the spin, to the molecule. This conclusion, however, does not hold when a more accurate profile, like $E^{\text{foc}}_M(\bm{R})$, is taken into account. As revealed by Eq.~\eqref{eq_gradient} with $M=0$, even a non-vortex Gaussian beam has a small impact on molecular rotations, in agreement with Ref.~\cite{Babiker_2002}. This physical behavior has a simple explanation. For instance, in the case of a diatomic molecule, the two nuclei, in general, experience a slightly different magnitude of the electric field. Within the electric-quadrupole interaction, this spatial gradient induces the OAM transfer, as suggested by Eq.~\eqref{eq_Hamiltonian_quad}.

\bibliography{main}

\end{document}